\documentclass[]{raa}            
\usepackage{graphicx,times}
\usepackage{natbib}

\begin{document}


\newcommand{\beq}{\begin{equation}}
\newcommand{\eeq}{\end{equation}}
\newcommand{\bd}{\begin{displaymath}}
\newcommand{\ed}{\end{displaymath}}
\newcommand{\bea}{\begin{eqnarray}}
\newcommand{\eea}{\end{eqnarray}}
\newcommand{\vect}[1]{\ensuremath{\mathbf{#1}}}
\newcommand{\mum}{\ensuremath{\, \mu}\mathrm{m}}
\newcommand{\m}{\mathrm{\, m}}
\newcommand{\M}{\mathrm{\, M}}
\newcommand{\mm}{\mathrm{\, mm}}
\newcommand{\cm}{\mathrm{\, cm}}
\newcommand{\nm}{\mathrm{\, nm}}
\newcommand{\km}{\mathrm{\, km}}
\newcommand{\AU}{\mathrm{\, AU}}
\newcommand{\g}{\mathrm{\, g}}
\newcommand{\s}{\mathrm{\, s}}
\newcommand{\pc}{\mathrm{\, pc}}
\newcommand{\yr}{\mathrm{\, yr}}
\newcommand{\Myr}{\mathrm{\, Myr}}   
\newcommand{\Gyr}{\mathrm{\, Gyr}}   
\newcommand{\Jy}{\mathrm{\, Jy}}
\newcommand{\mJy}{\mathrm{\, mJy}}
\newcommand{\erg}{\mathrm{\, erg}}
\newcommand{\K}{\mathrm{\, K}}
\newcommand{\D}{\mathrm{\, d}}


   \title
   {
     Debris Disks: Seeing Dust, Thinking of Planetesimals and Planets
   }

   \volnopage{ {\bf 2010} Vol.\ {\bf X} No. {\bf XX}, 000--000}
   \setcounter{page}{1}

   \author
   {
     A. V. Krivov
     \inst{}
   }

   \institute
   { 
     Astrophysical Institute and University Observatory,
     Friedrich Schiller University Jena,\\
     Schillerg\"a{\ss}chen 2--3, 07745 Jena, Germany;
     {\it krivov@astro.uni-jena.de}\\
     \vs \no
     {\small Received 2010 March 5; accepted 2010 March 24 }
   }

\abstract
{
Debris disks are optically thin, almost gas-free dusty disks observed
around a significant fraction of main-sequence stars older than about
10 Myr. Since the circumstellar dust is short-lived, the very existence of
these disks is considered as evidence that dust-producing planetesimals are still present
in mature systems, in which planets have formed~--- or failed to form~--- a
long time ago. It is inferred that these planetesimals orbit their host
stars at asteroid to Kuiper-belt distances and continually supply
fresh dust through mutual collisions.
This review outlines observational techniques and results on debris disks,
summarizes their essential physics and theoretical models,
and then places them into the general context of planetary systems,
uncovering
interrelations between the disks, dust parent bodies, and planets.
It will be shown that debris disks
can serve as tracers of planetesimals and planets
and shed light on the 
planetesimal and planet formation processes that operated in these systems in the past.
}

   \authorrunning{A. V. Krivov}  
   \titlerunning{Debris disks in planetary systems}   
   \maketitle


\section{Introduction}
\label{sect:intro}

An inventory of our own planetary system uncovers its complex architecture.
Eight known planets are arranged in two groups, four terrestrial ones
and four giants.
The main asteroid belt between two groups of planets, terrestrial and giant ones,
comprises planetesimals that failed to grow to planets because of the strong 
perturbations
by the nearby Jupiter \citep[e.g.,][]{safronov-1969,wetherill-1980}.
The Edgeworth-Kuiper Belt (EKB) exterior
to the Neptune orbit is built up by planetesimals that did not form planets
because the density of the outer solar nebula was too low
\citep[e.g.,][]{safronov-1969,lissauer-1987,kenyon-bromley-2008}.
Both the asteroid belt and the Kuiper belt are heavily sculptured by planets,
predominantly by Jupiter and Neptune respectively.
Accordingly, they include non-resonant and resonant families,
as well as various objects in transient orbits ranging from detached and scattered
Kuiper-belt objects through Centaurs to Sun-grazers.
Short-period comets, another tangible population of small bodies in the inner
solar system, must be genetically related to the Kuiper belt that act as their
reservoir \citep{Quinn-et-al-1990}.
Asteroids and short-period comets together are sources of interplanetary
dust, observed in the planetary region,
although their relative contribution to the dust production remains uncertain
\citep{Gruen-et-al-2001}.
And this complex system structure was likely quite different in the past.
It is argued that the giant planets and the Kuiper belt have originally formed in
a more compact configuration (the ``Nice model'',
\citeauthor{Gomes-et-al-2005} \citeyear{Gomes-et-al-2005}), and that
it went through a short-lasting period of dynamical instability, likely
explaining the geologically recorded event of the Late Heavy Bombardment (LHB).

Similar to the solar system, planetary systems around other stars are more
than a star itself and one or several planets.
As \citet{wyatt-2008} pointed out, at the end of the protoplanetary phase
a star is expected to be surrounded by one or all of the following components:
various planets from sub-Earth to super-Jupiter size; remnant of the protoplanetary
disk, both dust and gas; planetesimal belts in which solids continue to grow;
planetesimal belts, which are being ground down to dust.  
And indeed, many systems comprise
numerous smaller objects, ranging in size presumably from dwarf planets (like Pluto 
or Ceres)
down to dust. Evidence for this comes from
observations of the thermal emission and stellar light scattered by that dust.
A common umbrella term for all these ``sub-planetary'' solids
is a ``debris disk''.
Debris disks are aftermath of planet formation in the past, and they
formed in the same process as the planets did. Even in mature systems
where the planet formation has long been completed, they continue to evolve
collisionally and dynamically, are gravitationally sculptured by planets,
and the dust they produce through ongoing collisional cascades
responds sensitively
to electromagnetic and corpuscular radiation of the central star.
Hence debris disks can serve as indicators of directly invisible small bodies,
planets, and even stars, are tracers of their formation and evolution,
and represent an important component of planetary systems.

We start with outlining observational techniques and results on debris disks 
(Section~\ref{sect:observations}) and
summarizing their essential physics (Section~\ref{sect:physics})
and models (Section~\ref{sect:models}).
In the subsequent sections, the observational data and their theoretical
interpretation are used to draw conclusions about the ``dust end''
of the size distribution, dust itself 
(Section~\ref{sect:dust}),
its parent planetesimals
(Section~\ref{sect:planetesimals}),
planets expected to be present in debris disks systems
(Section~\ref{sect:planets}),
and the entire planetary systems
(Section~\ref{sect:planetary_systems}).
Section~\ref{sect:conclusions} provides a summary and lists open questions
of the debris disk research.

\section{Observational methods}
\label{sect:observations}

\subsection{Infrared Excesses}
An efficient, and historically the first successful, way of detecting 
circumstellar dust is the infrared (IR) photometry. If dust is present around
a star, it comes to a thermal equilibrium with the stellar radiation.
With equilibrium temperatures of dust orbiting a solar-type star at several tens
of AU being typically several tens of Kelvin, dust re-emits
the absorbed stellar light at wavelengths of several tens of micrometers,
i.e. in the far-IR.
As a result, the dust far-IR emission flux may
exceed the stellar photospheric flux at the same wavelength
by two-three orders of magnitude.
The first discovery of the excess IR emission around a main-sequence star,
namely Vega, was done by the Infrared Astronomical Satellite (IRAS)
\citep{aumann-et-al-1984}. Since then, IR surveys with IRAS, ISO, Spitzer,
and other space-based and ground-based telescopes have shown ``the Vega phenomenon''
to be common for  main-sequence stars.
In particular,
various photometric surveys of nearby stars have
been conducted with the Spitzer Space Telescope.
These are the GTO survey of FGK stars
\citep{beichman-et-al-2005,bryden-et-al-2006,beichman-et-al-2006b},
the FEPS Legacy project
\citep{meyer-et-al-2004,kim-et-al-2005,hillenbrand-et-al-2008,carpenter-et-al-2009},
the A star GTO programs \citep{rieke-et-al-2005,su-et-al-2006},
young cluster programs \citep{gorlova-et-al-2006,siegler-et-al-2006},
binary star programs \citep{trilling-et-al-2007},
as well as their combinations \citep[e.g.][]{trilling-et-al-2008}.
These observations were done mostly at
$24$ and $70\mum$ with the MIPS photometer
and resulted in hundreds of detections.
At MIPS sensitivity level, the incidences of debris disks
around different-ages stars with spectral classes from A to K
is about 15\%
\citep{su-et-al-2006,siegler-et-al-2006,trilling-et-al-2008,hillenbrand-et-al-2008}.   
The PACS instrument aboard the Herschel Space Observatory is now performing
observations (photometry and low-resolution ``small maps'') at a much
higher sensitivity level than that of Spitzer/MIPS.
The sub-mm wavelength region is being probed by CSO/SHARC, Herschel/SPIRE,
JCMT/SCUBA, APEX/LABOCA, and other instruments.
Soon more high-quality data should start coming from JCMT/SCUBA-2
\citep{matthews-et-al-2007}.

\subsection{Imaging}
So far, about 20 brightest debris disks have been resolved at various wavelengths
from visual through mid- and far-IR to sub-mm and radio;
see, e.g., http://www.circumstellardisks.org.
Scattered light images in the visual and near-IR are being taken, for instance,
with HST/ASC and many ground-based instruments in the coronagraphic mode,
thermal emission images in the mid-IR with Gemini South/TReCS and KeckII/MIRLIN,
whereas the JCMT/SCUBA instrument has been particularly successful in resolving debris disks
in the sub-mm.
Resolved systems span ages from about $12\Myr$ for the members of
the Beta Pictoris Moving Group
\citep{zuckerman-et-al-2001,ortega-et-al-2002}
such as $\beta$~Pic itself \citep{smith-terrile-1984}
or AU~Mic \citep{kalas-et-al-2004}
to $\sim 4$--$12\Gyr$ \citep{difolco-et-al-2004} for $\tau$~Cet \citep{greaves-et-al-2004}.
All of the disks resolved so far exhibit various structures: rings, inner voids,
clumps, spirals, warps, wing asymmetries.
An interpretation of these structures is dicsussed in Sect.~\ref{sect:planets}.

\subsection{Infrared Spectroscopy}
Early successful observations of several debris disks were made with NTT/TIMMI2
\citep{schuetz-et-al-2005}.
However, most of the information comes from the Spitzer/IRS instrument that have 
taken good-quality spectra of more than a hundred of stars from $5$ to $35\mum$.
For instance, \citet{chen-et-al-2006} describe IRS spectra of 59 stars with IRAS
$60\mum$ IR excesses, five of which show a clear silicate feature at $10\mum$.
Often other spectral features are seen that may be indicative of Fe-rich sulfides, water ice,
and other compounds \citep[e.g.][]{lisse-et-al-2007}.
In addition, another Spitzer instrument, MIPS, offered the so-called
spectral energy distribution (SED) mode, in which low-resolution spectra in the
region from $55\mum$ to $90\mum$ have been taken \citep{chen-et-al-2008}.
In the future, high-quality mid-IR spectra are expected, e.g., from JWST/MIRI.

\subsection{Other Methods}
One method is near-IR polarimetry. A certain degree of linear polarization of scattered light
coming from circumstellar debris dust is expected, because it has a flattened, disk geometry
\citep[e.g.][]{krivova-et-al-2000}.
Successful polarimetry observations were done so far
only for a few brightest debris disks,
such as $\beta$~Pic \citep{gledhill-et-al-1991,wolstencroft-et-al-1995}
and AU~Mic \citep{graham-et-al-2007}.
Such measurements can be particularly useful to further constrain the size distribution
and chemical composition of dust \citep{krivova-et-al-2000}.
Another method  is near-IR interferometry, which
probes dust in extreme proximity to the stars
(the so-called ``exozodiacal clouds'').
Using this method, a handful of  exozodis within  1~AU from the stars
was discovered in the recent years with the CHARA/FLUOR instrument in the K band at $2.2\mum$
\citep{absil-et-al-2006,difolco-et-al-2007,absil-et-al-2008,akeson-et-al-2009}
and with the Keck Interferometer Nuller at $10\mum$
\citep{stark-et-al-2009}.
An example of more exotic methods suggested to detect debris disks is microlensing
\citep{heng-keeton-2009}.

\section{Basic physics}
\label{sect:physics}

Every debris disk is composed of solids, spanning a
huge size range from hundreds of kilometers (large planetesimals) down
to a fraction of a micrometer (fine dust). These objects orbit the central
star and occasionally collide with each other.
We start with a brief
characterization of their individual orbits and then move on to the mutual collisions.

\subsection{Stellar Gravity and Radiation Pressure}
The force keeping a planetesimal or a dust grain on a closed orbit
is the central star's gravity:
\beq
  \vect{F}_{\rm g} = - \frac{GM_\star m}{r^3} \vect{r},
\eeq
where $G$ is the gravitational constant, $M_\star$ the stellar mass,
${\vect r}$ the radius vector of the object and $m$ its mass.
For the time being, we neglect further possible
gravitational forces by planets in the system (see 
Sect.~\ref{sect:planets}
for a discussion of their effects).
At dust sizes ($s \la 1\mm$), the solids are also subject to radiation pressure 
caused by the central star. Like stellar gravity, radiation pressure scales as
a reciprocal of the square distance to the star, but is directed outward.
Thus the two forces can be combined into a 
``photogravitational'' force \citep{burns-et-al-1979} that can be written as
\beq
  \vect{F}_{\rm pg} = -\frac{G M_\star (1-\beta) m }{r^3}\vect{r}.
\label{F_pg}
\eeq
Here, $\beta$ is the radiation pressure to gravity ratio that
depends on the grain size and optical properties.
The simplest assumption is that of compact grains of spherical shape that are 
characterized by their radius
$s$, bulk density $\rho$, and radiation pressure efficiency $Q_{\rm pr}$. The last
coefficient controls the fraction of the momentum transferred from the infalling
radiation to the grain. Its values range from 0 for perfect
transmitters to 2 for perfect backscatterers. An ideal absorber's radiation
pressure efficiency equals unity. In that case, the
$\beta$-ratio is given by \citep{burns-et-al-1979}
\bea
  \beta
  &=&
  0.574
  \left( \frac {L_\star} {L_\odot} \right)
  \left( \frac {M_\odot} {M_\star} \right)
  \left( \frac {1 \g \cm^{-3}} {\rho} \right)
  \left(\frac{ 1\:\mum}{s} \right),
\eea
where $L/L_\odot$ and $M/M_\odot$ are luminosity and mass of the star in solar
units.

\begin{figure}[h!!!]
   \centering
   \includegraphics[width=0.45\textwidth, angle=0]{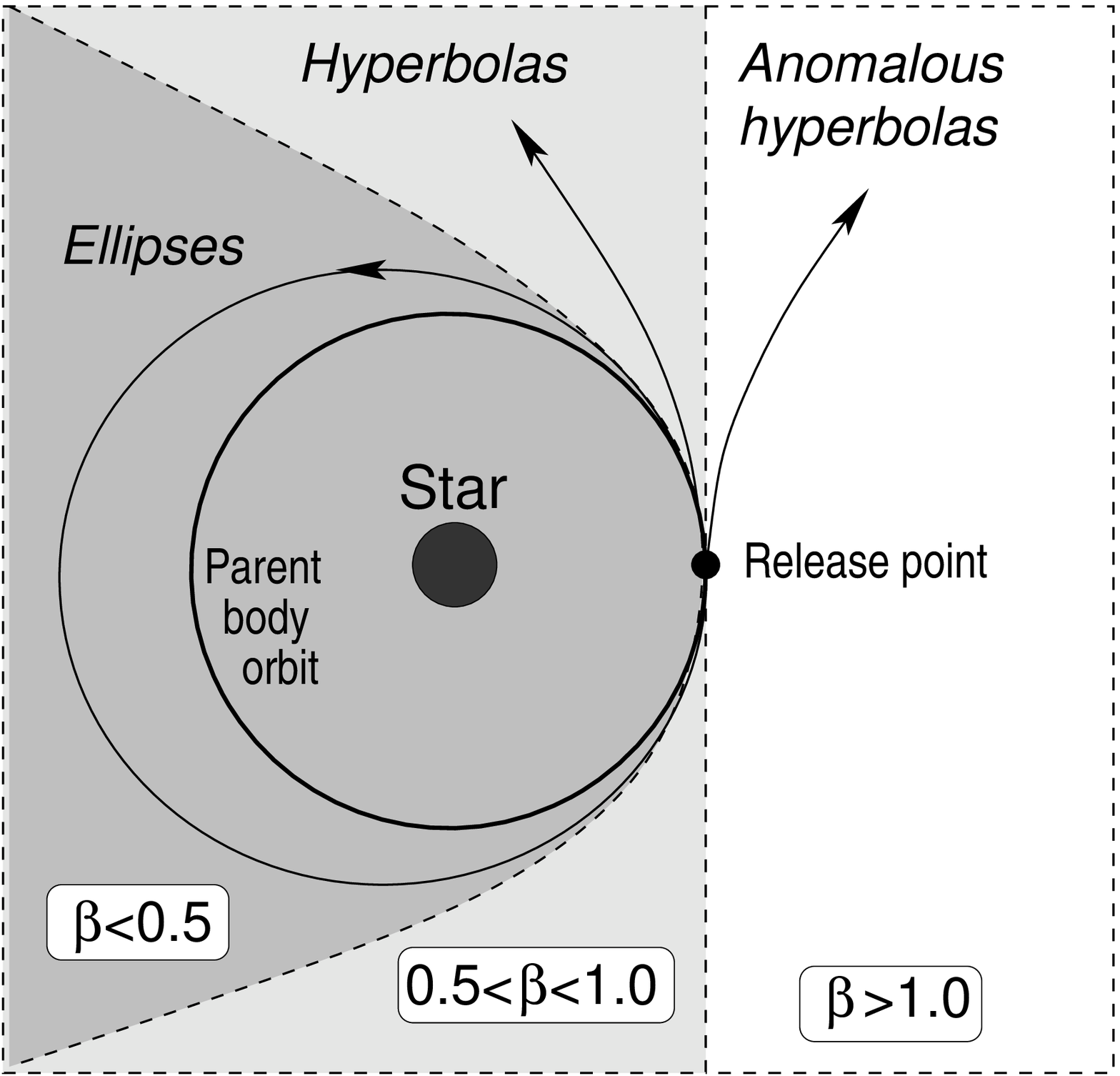}
   \includegraphics[width=0.45\textwidth, angle=0]{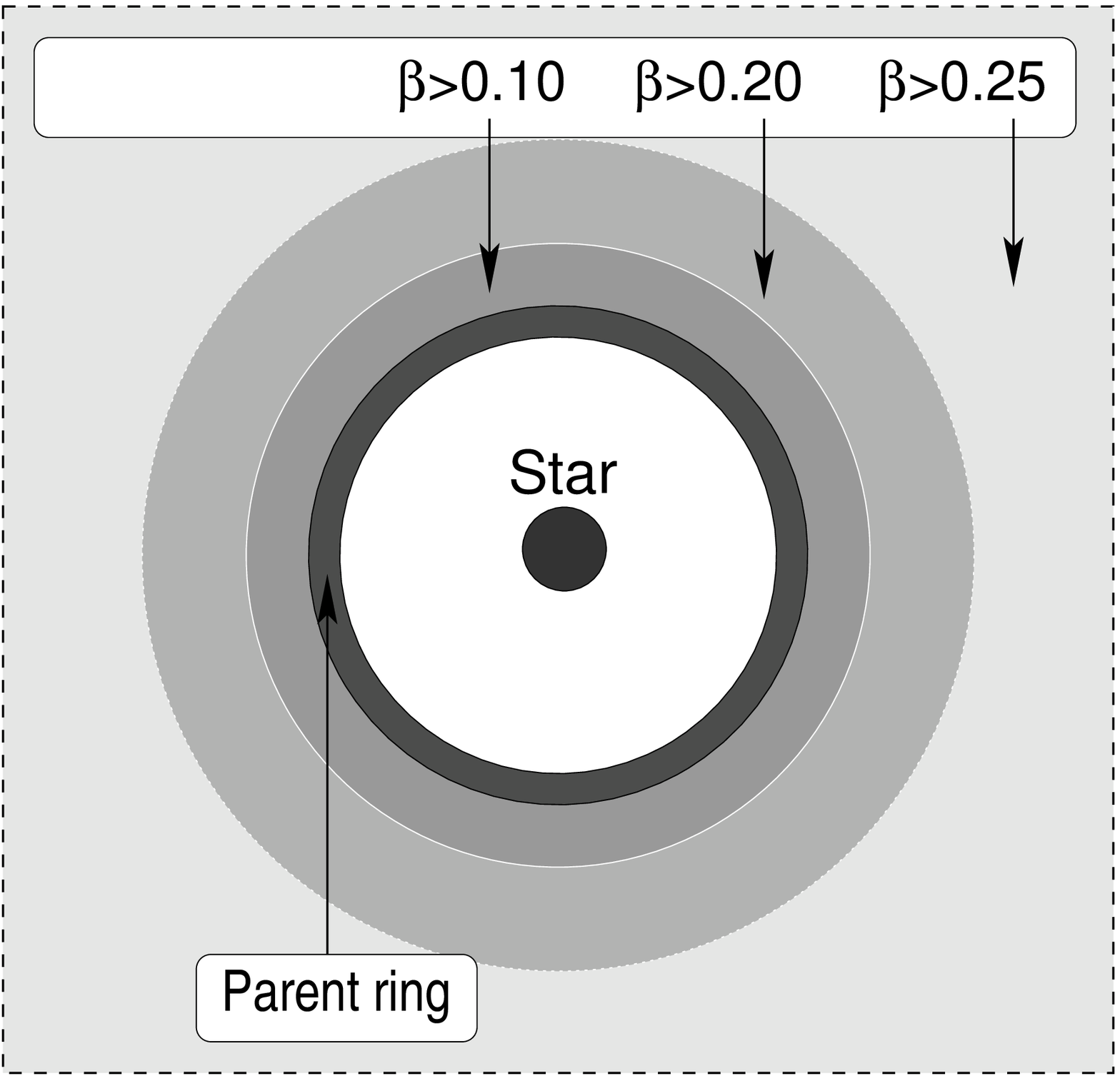}
   \begin{minipage}[]{1.0\textwidth}
   \caption{
   {\it Left:}
   Three possible types of orbits of dust particles under the combined action
   of stellar gravity and direct radiation pressure. For illustrative purposes,
   grains are assumed to be released from a circular orbit.
   {\it Right:}
   Schematic of a debris disk produced by a parent planetesimal belt.
   \label{fig:typical_disk}
   }
   \end{minipage}
   \end{figure}

If, by fragmentation or any other erosive process, smaller particles are released 
from larger ones,
they start to ``feel'' a photogravitational force rather than gravitational force.
The smaller the grains, the more the radiation
pressure they experience compensates the central star's gravity.
Thus the orbits of small particles differ from those of the parent bodies
(Fig.~\ref{fig:typical_disk} left).
A parent body on a circular orbit, for example,
releases fragments into bound elliptic orbits with
larger semimajor axes and eccentricities up to $\beta = 0.5$.
If $0.5 < \beta < 1$,
the grain orbits are hyperbolic and thus unbound.
For parent bodies in elliptic orbits, the boundaries
between bound and unbound orbits are somewhat smeared,
because the type of orbit depends on the ejection point.
All grains with $\beta < 1$ orbit the star on Keplerian trajectories at 
velocities reduced by a factor of $\sqrt{1 - \beta}$ compared to macroscopic bodies,
which are purely under the influence of gravitation.
Finally, below the critical size where $\beta = 1$ the effective force is repelling,
and the grains move on ``anomalous''(i.e., open outward) hyperbolic orbits.
Borrowing the terminology from solar system studies \citep{zook-berg-1975},
we may call dust grains in bound and unbound orbits
$\alpha$-meteoroids and $\beta$-meteoroids, respectively.

\subsection{Collisions}
As the very name ``debris disks'' suggests,
they should incessantly produce ``collisional debris'', so that
destructive collisions must be a 
dominant process operating in these systems.
For collisions to be destructive~--- and even to occur
at sufficient rates~-- a certain minimum level of relative velocities
is necessary. In protoplanetary disks, this condition is not fulfilled,
as relative velocities are strongly damped by dense ambient gas, so that the
orbits all have low eccentricities and inclinations.
In contrast, debris disks are gas-poor, and damping is not efficient at all.
However, the absence of damping is necessary, but not sufficient 
for relative velocities to be high.
As a protoplanetary disk transforms to a debris disk in the course of its
evolution, solids are expected to preserve a low velocity dispersion they had 
before the primordial gas removal. Accordingly, to ignite destructive 
collisions, debris disks must be ``stirred'' by some mechanism
\citep{wyatt-2008}.
This can be self-stirring by largest planetesimals
\citep{kenyon-bromley-2004a} or stirring by planets orbiting in the inner
gap of the disk \citep{wyatt-2005b,mustill-wyatt-2009}.
Further possibilities which, however, are not considered typical for the majority
of the disks, include ``fast ignition'' by 
the abrupt injection of  a planet (suggestively of Neptune masses) into the 
disk after planet-planet scattering \citep{raymond-et-al-2009}
or external events such as stellar flybys \citep{kenyon-bromley-2002}.

Once the disk is sufficiently stirred, the collisional cascade sets in.
The material is ground all the way down to the dust sizes, until the smallest
fragments with $\beta \ga 0.5$ are blown away
from the disk by radiation pressure.
Note that at dust sizes, stirring is no longer required. As typical
eccentricities of grains are of the same order as their $\beta$-ratio
\citep[e.g.][]{burns-et-al-1979}, radiation pressure ensures the impact velocities
to be sufficiently high. What is more, some of the hyperbolic grains,
$\beta$-meteoroids, may be energetic enough to trigger a self-sustained
cascade (``dust avalanche'') on their way outward through the disk
\citep{artymowicz-1997,krivov-et-al-2000b}.
This process, however, can only be efficient in the dustiest debris disks
such as $\beta$~Pic \citep{grigorieva-et-al-2006}.

We now take a more quantitative look at the collisions.
An important quantity in the collisional prescription is the critical
specific energy for disruption and dispersal, $Q_{\rm D}^\star$.
It is defined as the impact energy per unit target mass that results in
the largest remnant containing a half of the original target mass.
For small objects, $Q_{\rm D}^\star$ is determined solely by the material
strength, while for objects larger than $\sim 100\,\m$, the gravitational
binding energy dominates.
As a result, $Q_{\rm D}^\star$ is commonly described by the sum of two
power laws \citep[see, e.g.,][]{davis-et-al-1985,holsapple-1994,%
paolicchi-et-al-1996,durda-dermott-1997,durda-et-al-1998,%
benz-asphaug-1999,kenyon-bromley-2004b,stewart-leinhardt-2009}:
\begin{equation}
  Q_{\rm D}^\star
   = Q_{\rm s}
     \left(\frac{s}{1~\mathrm{m}}\right)^{-b_{\rm s}}
   + Q_{\rm g}
     \left(\frac{s}{1~\mathrm{km}}\right)^{b_{\rm g}},
\label{eq:QD}
\end{equation}
where the subscripts s and g stand for the strength and gravity regime,
respectively. The values of $Q_{\rm s}$ and $Q_{\rm g}$ lie in 
the range $\sim 10^5$--$10^7\erg\g^{-1}$,
$b_{\rm s}$ is between 0 and 0.5,
and $b_{\rm g}$ between 1 and 2
\citep{benz-asphaug-1999}.
The critical energy reaches a minimum
($\sim 10^4$--$10^6\erg\g^{-1}$) at sub-km sizes.
At dust sizes, Eq.~(\ref{eq:QD}) suggests $ \sim 10^8\erg\g^{-1}$,
but true values remain unknown, as this size regime has never been probed 
experimentally.
Both colliders are disrupted if their relative velocity $v_\mathrm{rel}$ exceeds
\citep[see, e.g.][their Eq. 5.2]{krivov-et-al-2005}
\be
  v_{\mathrm{cr}} = \sqrt{
    {2 (m_\mathrm{t} + m_\mathrm{p})^2  \over m_\mathrm{t} m_\mathrm{p}}
    Q_{\rm D}^\star
                         } ,
\label{disruption}
\ee
where
$m_\mathrm{t}$ and $m_\mathrm{p}$ are masses of the two colliders.
Thus two like-sized objects will be destroyed if they collide
at a speed exceeding $\sqrt{8Q_{\rm D}^\star}$.
For two dust grains, the critical speed is several hundreds 
$\m\s^{-1}$. At several tens of AU from a star, this would imply
eccentricities of $\sim 0.1$ or higher. Such eccentricities can be
naturally attained by dust grains with $\beta \sim 0.1$ as a result of
the radiation pressure effect described above. 

The actual physics of collisions in debris disks is
more complicated. 
Collisions occur across the size range that is $\sim 12$ orders of magnitude wide
and at quite different velocities.
Statistically, collisions between objects of {\it very} dissimilar sizes
(e.g., impacts of dust grains onto ``boulders'') outnumber those
between like-sized objects.
Besides, grazing or at least oblique collisions are more frequent than
frontal ones.
All this leads to a spectrum of possible collisional outcomes
that take place simultaneously in the same disk, comprising further growth of large
planetesimals, partial destruction (cratering) of one or both projectiles
with or without reaccumulation of fragments,
as well as a complete disruption of small planetesimals and dust particles.
Detailed analyses show that two outcomes are expected to play a crucial role
in debris disks:
complete disruption and cratering \citep{thebault-augereau-2007,mueller-et-al-2009}.

\subsection{Drag Forces}
A more accurate, special-relativistic, derivation of the photogravitational force
leads to an additional velocity-dependent term that has to be added
in the right-hand side of Eq.~(\ref{F_pg}) \citep{burns-et-al-1979}:
\beq
  {\bf F}_{\rm PR} = - \frac{G M_\star \beta m }{r^2}
  \left[
    \left(   { {\bf v}{\bf r} \over c r} \right) { {\bf r} \over r }
           + { {\bf v} \over c }
  \right]  
 \label {F_PR}
\eeq
with ${\bf v}$ being the velocity vector of the particle,
which is referred to as the Poynting-Robertson (P-R) force.
Being dissipative, this force causes a
particle to lose gradually its orbital energy and angular momentum.
Thus all $\alpha$-meteoroids move in Keplerian ellipses with reducing
semimajor axes $a$ and eccentricities $e$ \citep{wyatt-whipple-1950}
(except very close to the central star,
\citeauthor{breiter-jackson-1998}
\citeyear{breiter-jackson-1998}).
On timescales of thousands of years or more the trajectory shrinks to the
star.

Like stellar electromagnetic radiation gives rise to radiation pressure
forces, stellar particulate radiation~--- stellar wind~--- causes stellar wind
forces. Similarly to the net radiation pressure force, the total stellar
wind force can be decomposed to direct stellar wind pressure and 
stellar wind drag. 
For most of the stars, the momentum and energy flux carried by
the stellar wind is by several orders of magnitude smaller than that carried by
stellar photons, so that the direct stellar wind pressure is negligibly small.
However, the stellar wind drag forces cannot be ignored, because the stellar wind velocity
$v_{\rm sw}$, to replace $c$ in Eq.~(\ref{F_PR}), is much smaller than $c$.
Stellar wind forces can be very important in debris disks
around late-type stars 
\citep{plavchan-et-al-2005,strubbe-chiang-2006,Augereau-Beust-2006}.  

Compared to gas-rich protoplanetary disks, debris disks are expected to be
extremely gas-poor, because primordial gas disperses on timescales
$\la 10\Myr$ \citep{alexander-2008,hillenbrand-2008}.
Even so, in youngest debris disks with ages of $10$--$30\Myr$,
little amounts of gas have been detected.
Debris disk gas might be either the remnant of the primordial protoplanetary disk
or, more likely, of secondary origin.
Possible mechanisms for producing secondary gas include
photon-induced desorption from solids
\citep{chen-et-al-2007} and grain-grain collisions
\citep{czechowski-mann-2007}.
Part of the observed gas may also stem from
comet evaporation, as inferred from observed time-variable stellar absorption lines
\citep[e.g.][]{ferlet-et-al-1987,beust-valiron-2007}.
In some systems, exemplified by $\beta$~Pic \citep{olofsson-et-al-2001} and 
HD~32297 \citep{redfield-2007}, evidence for a 
stable, orbiting gas component was found.
Dynamical effects exerted by rotating gas on dust in young debris disks
have been studied, for instance, by \citet{thebault-augereau-2005}
and \citet{krivov-et-al-2009}. 

\subsection{Other Forces and Effects}
Dust grains interacting with the stellar radiation and stellar wind
acquire electrostatic charges due to a variety of effects (photoelectron emission, 
secondary electron emission, sticking of electrons and ions etc.). Under the 
presence of the star's magnetic field, they experience the Lorentz force.
Its influence on the dynamics of interplanetary dust
particles in the solar system is well understood.
As the dust grains move through the sectored magnetic field of the Sun
with alternative polarities, the Lorentz force rapidly changes its
direction. The mutual near-cancellation of these contributions is not
complete, however. On long time spans the Lorentz force results in   
a ``Lorentz diffusion'', i.e. essentially stochastic changes in $a$, $e$  
and, most importantly, inclination $i$
\citep{morfill-gruen-1979a,consolmagno-1980,%
barge-et-al-1982b,wallis-hassan-1985}.
The effect becomes significant for grains less than several micrometers in
size.
The Lorentz force is not included in most of the debris disk models.
This is because it affects relatively small grains,
which make only a minor contribution both into 
the mass budget and the total cross section area of the dust disk.

Sublimation, or transition of dust grains from the solid to the vapor state,
occurs in the vicinity of the star. The radius of the sublimation zone depends
strongly on the material and porosity of the dust grains and on the luminosity
of the primary.
In the solar system, silicate and carbonaceous grains sublimate at 2--4 solar radii
from the Sun \citep[e.g.][]{kimura-et-al-1997,krivov-et-al-1998b}.
However, for a water ice~-- organics mixture the sublimation distance in the solar 
system is as large as $\sim 20\AU$, and increases to $30$-$40\AU$ around
luminous A-stars \citep{kobayashi-et-al-2009}.

A number of more exotic effects exist that may influence the dust particles 
themselves and their dynamics \citep{burns-et-al-1979,kapisinsky-1984}.
Some are related to the fact that the star is not a point-like object
(differential Doppler effect, radiation pressure from an extended source).
Others stem from stellar radiation together with rotation of objects
(Yarkovsky effect, windmill effect, Radzievsky effect).
Still others are related to modification of the grain properties by the
environment (sputtering by plasma or interstellar grains, packing effect,
electrostatic breakup), etc.
None of these, however, seem to be of generic importance for debris disks,
requiring either closeness to the star, smallness of particles, or special
material compositions.

\section{Models}
\label{sect:models}

\subsection{Numerical Simulation Methods}
As described in Sect.~\ref{sect:physics},
each debris disk can be treated an ensemble of objects of different sizes
(from dust to planetesimals)
that orbit the star under the influence of gravity, radiation pressure,
and other forces.
In addition, they experience collisions which destroy or erode these
objects and produce new ones, collisional fragments.
Such a system can be numerically modeled by a variety of methods
\cite[see][for a summary]{krivov-et-al-2005} that can be classified into three
major groups: $N$-body simulations, statistical approach, and hybrid methods.

{\it $N$-body codes} follow trajectories of individual of disk
objects by numerically integrating their equations of motion.
During numerical integrations, instantaneous positions and velocities of particles 
are stored. Assuming that the objects are produced and lost at constant rates,
this allows one to calculate a steady-state spatial distribution of particles
in the disk.
Then, collisional velocities and rates can be computed and collisions can be applied.
It is usually assumed that each pair of objects
that come in contact at a sufficiently high relative velocity is 
eliminated from the system without generating 
smaller fragments \citep{desetangs-et-al-1996b,stark-et-al-2009b,krivov-et-al-2009}
or produces a certain number of fragments of equal size \citep{wyatt-2006}.
The $N$-body method is able to handle an arbitrary large array of forces
and an arbitrary complex dynamical behavior of disk solids driven
by these forces.
Accordingly, it is superior to the others in studies of structures in debris
disks arising from interactions with planets, ambient gas, or interstellar medium.
However, this method cannot treat a large number of objects sufficient to cover
a broad range of particle masses and thus is less successful in
modeling the collisional cascade. In particular, an accurate 
characterization of the
size distribution with $N$-body codes is hardly possible.

{\it Statistical method} effectively replaces particles themselves with their distribution in 
an appropriate phase space. The ideas trace back to the classical calculation
of the velocity distribution in particle ensembles with Boltzmann equation
\citep{boltzmann-1896,chapman-cowling-1970} and
of the mass (or size) distribution with the coagulation
equation \citep{smoluchowski-1916,chandrasekhar-1943}.
Note that the term ``coagulation equation'' is actually used regardless of whether
the colliding particles merge, fragment, or get destroyed.
In applications to planetesimal and debris disks, one
introduces a mesh of several variables comprising, for instance,
mass, distance, and velocity
\citep{spaute-et-al-1991,weidenschilling-et-al-1997,%
kenyon-luu-1998,kenyon-luu-1999a,kenyon-luu-1999b,%
kenyon-bromley-2002,kenyon-bromley-2004a,kenyon-bromley-2004c,%
thebault-et-al-2003,thebault-augereau-2007}
or mass and orbital elements
\citep{delloro-paolicchi-1998,delloro-et-al-1998,delloro-et-al-2001,delloro-et-al-2002,%
krivov-et-al-2005}.
The number of particles in each bin of the mesh at successive time instants
is computed by solving equations that describe gain and loss of objects by collisions
and other physical processes.
Many of the results presented in this paper
(e.g., Fig.~\ref{fig:rad_size_dist})
were obtained with the code ACE (Analysis of Collisional Evolution)
that simulates the disks in such a manner, using a 3D mesh
of masses, pericentric distances, and eccentricities
\citep{krivov-et-al-2005,krivov-et-al-2006,krivov-et-al-2008}.
Statistical codes are much more accurate in handling collisions than $N$-body ones,
but treat dynamics in a simplified way, for instance, by averaging over
angular orbital elements.
For this reason, they are less suitable for
simulation of structures in debris disks.

Lastly, {\it hybrid codes}
\citep{weidenschilling-et-al-1997,goldreich-et-al-2004}
combine $N$-body integrations of a few large bodies
(planets, planetary embryos, biggest planetesimals) with a statistical simulation
of numerous small planetesimals and dust.
Such codes, originally developed for planetesimal accretion,
have also been applied to debris disks 
\citep[e.g.][]{bromley-kenyon-2006,kenyon-bromley-2008}.
Finally, a stand-alone ``superparticle'' method developed to study
major collisional break-ups in debris disks \citep{grigorieva-et-al-2006}
should be mentioned.
This method, too, combines $N$-body and statistical approaches, although
in a different way.

\subsection{A ``Standard Model'' for a Debris Disk}
A standard model of a debris disk
can be devised by considering two major effects,
stellar photogravity and collisions, and neglecting
the drag forces and all other processes
\citep[e.g.][]{thebault-et-al-2003,krivov-et-al-2006,strubbe-chiang-2006,%
thebault-augereau-2007,loehne-et-al-2007,krivov-et-al-2008,thebault-wu-2008}.
Imagine a relatively narrow belt of planetesimals (``birth ring'' or ``parent ring'')
in orbits with moderate eccentricities and inclinations, exemplified by the
classical Kuiper belt in the solar system (Fig.~\ref{fig:typical_disk} right).
The planetesimals orbiting in the birth ring undergo collisional cascade
that grinds the solids down to dust.
At smallest dust sizes, stellar radiation pressure effectively reduces
the mass of the central star and quickly (on the dynamical timescale)   
sends the grains into more eccentric orbits, with their pericenters still
residing within the birth ring while the apocenters are located outside
the ring.
As a result, the dust disk extends outward from the planetesimal belt.
The smaller the grains, the more extended their ``partial'' disk.
The tiniest dust grains, for which the radiation pressure effectively    
reduces the physical mass by half, are blown out of the system in
hyperbolic orbits.  
The radiation pressure blowout of the smallest collisional debris
represents the main mass loss channel in such a disk.

One expects a balance between the production of dust by the
collisional cascade and its losses by radiation pressure blowout.
If such a balance exists,
the amounts of particles with different sizes on different orbits
stay constant relative to each other, 
and a debris disk is said to be in a quasi-steady state.
However, the absolute amounts should adiabatically decrease with time,
because the material at the top-end of the size distribution is not
replenished (therefore ``quasi'').
For brevity, ``quasi'' is often omitted and a simple
``steady state'' is used.
Basic properties of steady-state disks, derived analytically
\citep[e.g.][]{strubbe-chiang-2006}
and by numerical simulations with statistical codes
\citep{krivov-et-al-2006,thebault-augereau-2007}
are described in subsequent subsections.
The steady-state evolutionary regime can be perturbed, for instance,
by occasional collisional break-ups of largest planetesimals or
by ``shake-downs'' of the system due to instability of nearby planets.
After such events, the disk needs some time to relax to a new steady state.

\subsection{Collision-Dominated and Transport-Dominated Disks}
The model just described is only valid, if the collisional timescale
is shorter than the characteristic timescale of the drag forces.
Such disks are usually referred to as collision-dominated, as opposed to
systems that~--- at dust sizes~--- are transport-dominated
\citep[e.g.][]{krivov-et-al-2000b}.
In the latter case, radial transport of dust material by various drag   
forces occurs on shorter timescales then collisions.
Then, additional removal mechanisms may play a significant role.
For example, Poynting-Robertson (P-R) drag can bring grains close to the
star where they would sublimate or deliver them into the planetary region
where they would be scattered by planets.
To be transport-dominated, the system should either have an optical depth
below current detection limits \citep{wyatt-2005} or be subject to
transport mechanisms other than P-R drag, such as strong stellar winds
around late-type stars 
\citep{plavchan-et-al-2005,strubbe-chiang-2006,Augereau-Beust-2006}.  
Most of debris disks detected so far are thought to be
collision-dominated.
Accordingly, transport-dominated systems will not be considered here.

Note that at sufficiently large sizes, all disks are dominated by collisions.
This is because the lifetime against catastrophic collisions in a disk
with the $\s^{-3.5}$ size distribution discussed below scales as
$\propto \sqrt{s}$ \citep[see, e.g., Appendix~A in][]{wyatt-et-al-1999},
whereas any drag force is proportional to the ratio of the mass and cross section,
i.e. $\propto s$.
For instance, the interplanetary dust cloud in the solar system at $1\AU$
from the Sun is tenuous enough to be P-R-dominated at dust sizes,
but is collision-dominated above $s \approx 100\mum$ \citep{gruen-et-al-1985}.

\subsection{Size and Radial Distribution in Collision-Dominated Disks}
We now consider typical size and radial distribution of material in
a steady-state collision-dominated debris disk, seen in simulations.
As an example, we take the ``reference model'' of the debris disk of Vega
from \citet{mueller-et-al-2009} that assumes a parent ring between $\approx 80$
and $\approx 120\AU$. However, the results are generic and
qualitatively hold for all collision-dominated disks,

   \begin{figure}[h!!!]
   \centering
   \includegraphics[width=0.43\textwidth, angle=0]{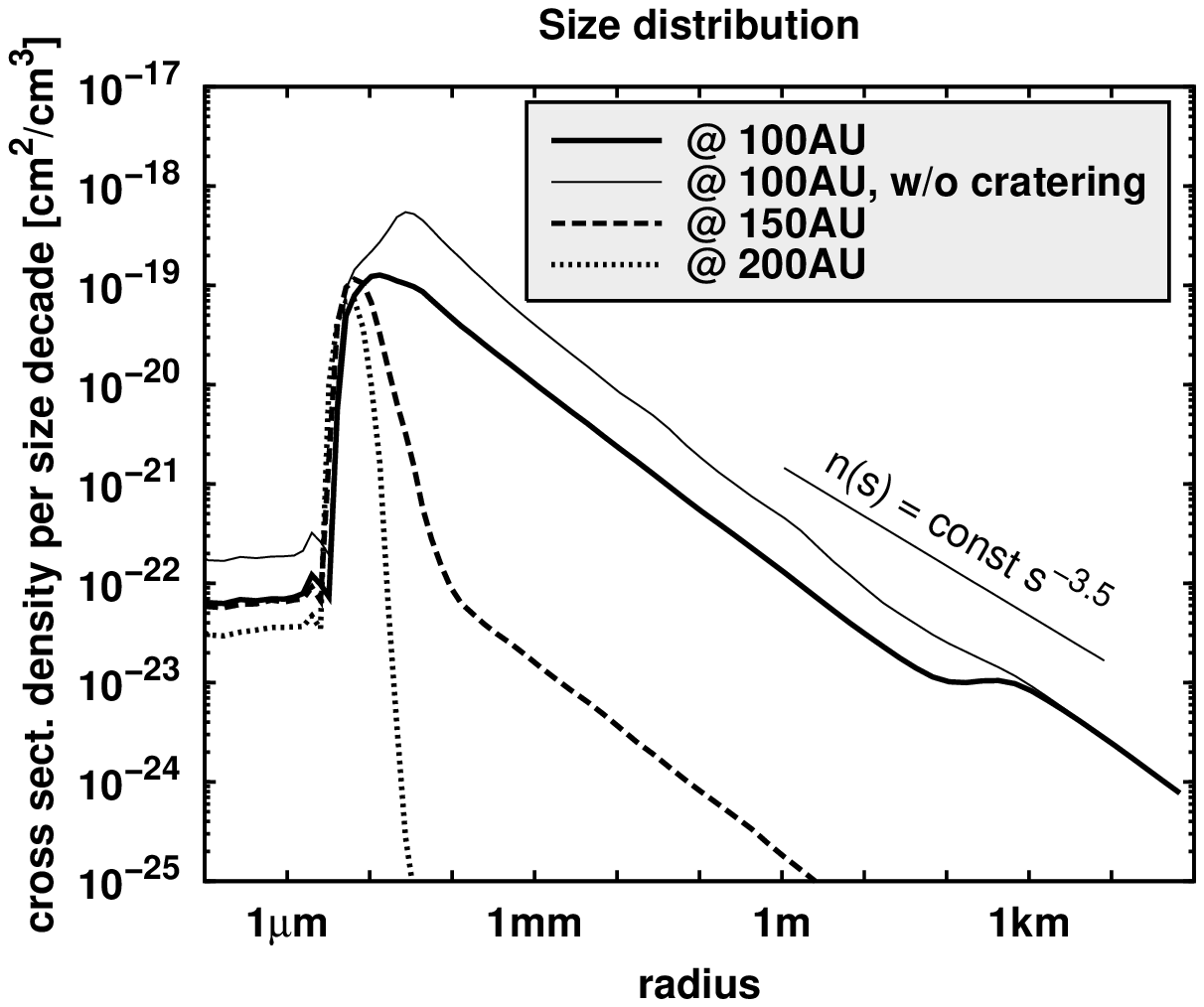}
   \hspace*{0.04\textwidth}
   \includegraphics[width=0.43\textwidth, angle=0]{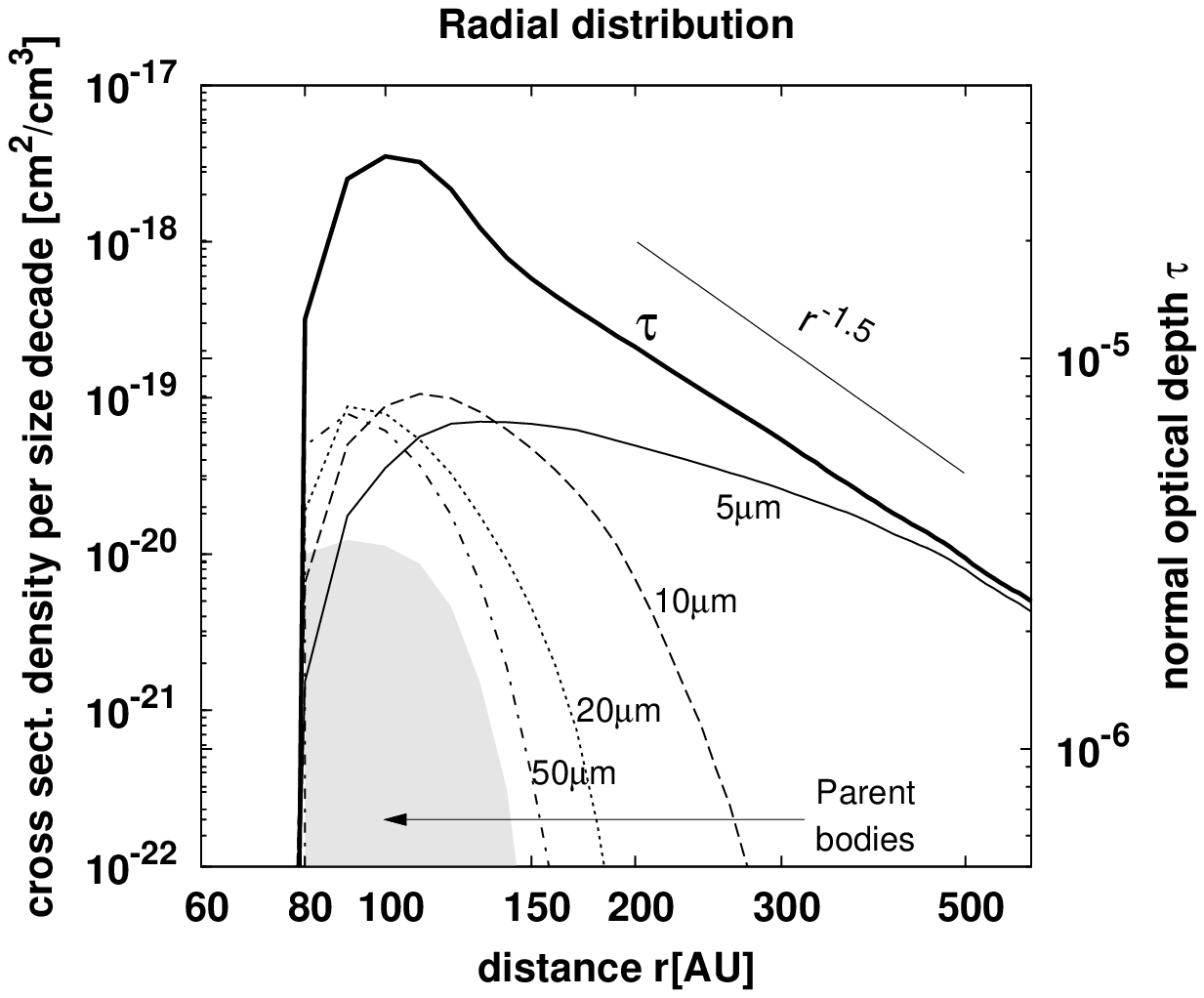}
   \begin{minipage}[]{1.0\textwidth}
   \caption{
   Modeled dust distributions in a collision-dominated debris disk, here:
   the ``reference'' model of \citet{mueller-et-al-2009}.
   {\it Left:}
   size distribution at different distances from the star with
   catastrophic and cratering collisions included in the model (thick lines).
   For comparison, a distribution at $100\AU$ computed without cratering
   is also shown (thin solid line).
   Straight line depicts the \citet{dohnanyi-1969} distribution for comparison.
   {\it Right:}
   radial distribution of different-sized grains (thin lines),
   as well as the overall radial profile of the normal geometrical optical depth
   (think solid line).
   Straight line is the $r^{-1.5}$ analytic prediction 
   for the optical depth \citep{strubbe-chiang-2006}.
   \label{fig:rad_size_dist}
   }
   \end{minipage}
   \end{figure}

Figure~\ref{fig:rad_size_dist} (left) depicts the size distribution.
Most noticeable is the peak at the particle size that corresponds to
$\beta \sim 0.5$, where the radiation pressure is half as strong as gravity.
For the Vega disk this blowout radius is $s_{\rm blow} \approx 4\mum$.
Below this size bound orbits are rare (and below the size
where $\beta = 1$ they are impossible), and the grains are blown away
on a disk-crossing timescale of the order of $10^2$ to $10^3$~yr.
Typically, the amount of blowout grains instantaneously present in the
steady-state system is much less than the amount of slightly larger
grains in loosely bound orbits around the star.
This is because the production rate of the grains of sizes 
slightly less and slightly greater than $s_{\rm blow}$ is
comparable, but the lifetime of bound grains (due to collisions) is much
longer than  the lifetime of blowout grains (disk-crossing timescale).
This explains a drop in the size distribution around the blowout size
seen in the left panel of Fig.~\ref{fig:rad_size_dist}.
The size distribution above $ s_{\rm blow}$ sensitively depends
on the collisional outcome model assumed, which is seen by comparing
the distributions at $100\AU$ computed with and without taking cratering collisions
into account.
The overall size distribution inside the parent ring (the $100\AU$ curve)
at sizes $s \gg s_{\rm blow}$ roughly follows
the classical Dohnanyi power law, which
corresponds to the differential size distribution
$n \propto s^{-3.5}$.
That law was theoretically derived by \citet{dohnanyi-1969}
for a collisional cascade in a system without sources and sinks,
composed of objects with $Q_{\rm D}^\star = {\rm const}$.
The deviations of the size distribution in Fig.~\ref{fig:rad_size_dist} (left)
from the Dohnanyi law reflect the size dependence of $Q_{\rm D}^\star$, given
by Eq.~(\ref{eq:QD}).
A notch at $s \sim 100\m$ corresponds to the minimum of $Q_{\rm D}^\star$,
and the slope at larger sizes is slightly steeper that Dohnanyi's, because
$Q_{\rm D}^\star$ increases with size in the gravity regime
\citep{durda-dermott-1997}.
Another generic feature of the size distribution is that it becomes narrower
at larger distances from the star. At twice the distance of the parent ring,
$200\AU$, the distribution converts to
narrow peak around $\sim 2 s_{\rm blow}$ composed only of small, high-$\beta$,   
barely bound grains sent by radiation pressure into eccentric orbits
with large apocentric distances.

The spatial distribution of dust in the same modeled disk is illustrated
by Figure~\ref{fig:rad_size_dist} (right).
As noted above, smaller grains with higher $\beta$ ratios acquire
higher orbital eccentricities.
Since the eccentricities of particles slightly above the blowout limit 
($s = 5\mum$) are the highest,
their radial distribution is the broadest, whereas larger particles
stay more confined to their birth regions.
In the figure,
this effect can be seen from how the curves gradually change from the smallest 
($s=5\mum$) to the largest bound grains ($50\mum$).
The latter essentially follow the distribution of the parent bodies.
In addition, the same figure shows
the radial profile of the normal geometrical optical depth.
The slope of $\tau \propto r^{-\alpha}$ is close to $\alpha = 1.5$,
as predicted analytically by \citet{strubbe-chiang-2006} for collision-dominated disks.

\subsection{Disk Appearance at Different Wavelengths}
\begin{figure}[h!!!]
   \vspace*{5mm}
   \centering
   \includegraphics[width=0.9\textwidth, angle=0]{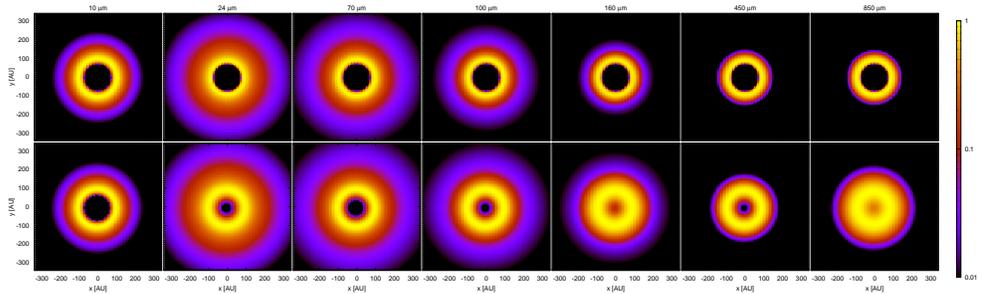}
   \begin{minipage}[]{1.0\textwidth}
   \caption{
   Appearance of a debris disk at different wavelengths from $10$ to $850\mum$.
   {\it Top row:}
   synthetic images of the Vega disk, based on the ``reference'' model of 
   \citet{mueller-et-al-2009}.
   {\it Bottom row:}
   the same images convolved with the Point-Spread Function (PSF), which
   we assumed to be a Gaussian with the width ($\sigma$) of
   $0.089''$ at $10\mum$ (Keck, mirror diameter $10\m$),
   $2.37''$ at $24\mum$ and $1.77''$ at $70\mum$ (Spitzer/MIPS, $0.9\m$),
   $2.53''$ at $100\mum$ and $4.05''$ at $160\mum$ (Herschel/PACS, $3.5\m$),
   $2.66''$ at $450\mum$ and $5.03''$ at $850\mum$ (JCMT/SCUBA, $15\m$).
   These convolved images show how the Vega disk would appear in observations with best
   present-day instruments operating at the respective wavelengths.
   Note that in actual sub-mm images the Vega disk appears clumpy
   \citep{holland-et-al-1998,marsh-et-al-2006}, which is not seen here,
   because the model by \citet{mueller-et-al-2009} is rotationally-symmetric.
   {\it Courtesy: Sebastian M\"uller.}
   \label{fig:syn_images}
   }
   \end{minipage}
   \end{figure}

Size segregation of dust in debris disks described above means
that in resolved images at different wavelengths the same disk would look
differently (Fig.~\ref{fig:syn_images}).
Measurements at longer wavelengths (sub-mm) probe larger grains, because
they are cooler (see Sect.~\ref{sect:dust}),
and thus trace the parent ring.
Such observations may also reveal clumps, if for instance there is a
planet just interior to the inner rim of the parent ring, and parent
planetesimals and their debris are trapped in external resonances,
similar to Plutinos in 3:2 resonance with Neptune
\citep{wyatt-2006,krivov-et-al-2006b}.
At shorter wavelengths (far-IR, mid-IR), smaller (warmer) grains are
probed.
Thus the same disk appears much larger. The image may be featureless, even
if the parent ring is clumpy, because strong radiation pressure
\citep{wyatt-2006} and non-negligible relative velocities
\citep{krivov-et-al-2006b} would liberate small particles from resonant clumps.
As a result, they would form an extended disk, as described above,
regardless of whether their parent bodies are resonant or not.
Finally, at shortest wavelengths of thermal emission, only the hottest
closest-in grains are evident in the observations, and thus again
only the parent ring is seen.

\subsection{Scaling Laws for Evolution of Collision-Dominated Disks}
Some useful laws that describe quasi-steady state evolution
of collision-dominated disks have been found
\citep{wyatt-et-al-2007,loehne-et-al-2007, krivov-et-al-2008}.

{\it 1. Dependence of evolution on initial disk mass.}
Consider a narrow parent ring with initial mass $M(t=0) \equiv M_0$ at a distance $r$ from 
the star with age $t$.
Denote by $F(M_0,r,t)$ any quantity directly proportional
to the amount of disk material in any size regime, from dust grains to
planetesimals.
In other words, $F$ may equally stand for the total disk mass,
the mass of dust, its total cross section, etc.
There is a scaling rule \citep{loehne-et-al-2007}:
\be
       F(x M_0, r, t) = x F(M_0, r, x t) ,
\label{scaling1}
\ee
valid for any factor $x > 0$. This scaling is an {\it exact} property of
every disk of particles, provided these are produced, modified and lost in
binary collisions and not in any other physical processes.

\bigskip
{\it 2. Dependence of evolution on distance.}
Another scaling rule is the dependence of the evolution timescale on
the distance from the star \citep{wyatt-et-al-2007,loehne-et-al-2007}.
Then
\be
       F(M_0, x r, t) \approx F(M_0, r,  x^{-4.3} t) .
\label{scaling2}
\ee
Unlike Eq.~(\ref{scaling1}), this scaling is approximate.

\bigskip
{\it 3. Dust mass as a function of time.}
Finally, the third scaling rule found in  
\citet{loehne-et-al-2007}
is the power-law decay of the dust mass
\be
       F(M_0, r, x t) \approx x^{-\xi} F(M_0, r, t) ,
\label{scaling3}
\ee
where $\xi \approx 0.3 \ldots 0.4$.
This scaling is also approximate and,
unlike Eq.~(\ref{scaling1}) and Eq.~(\ref{scaling2}), only applies
to every quantity directly proportional to the amount of {\it dust}.
In this context, ``dust'' refers to all objects in the strength rather than gravity
regime, implying radii less than about 100 meters.
The scaling is sufficiently accurate for disks that are much older than
the collisional lifetime of these $100\m$-sized bodies.

\section{Debris disks: Seeing dust}
\label{sect:dust}

\subsection{Dust Temperature and Fractional Luminosity}
Observational data for most of the debris disks are pretty scarce and consist
only of a few photometric points indicative of the IR excess.
In many cases, these points can be fitted with a
spectral energy distribution (SED) of a blackbody with a single temperature.
Two useful quantities that can be derived from such observations are
the dust temperature $T_{\rm d}$ and the dust fractional luminosity $f_{\rm d}$,
defined as a ratio of the bolometric luminosities of the dust and the star.

Both  $T_{\rm d}$ and $f_{\rm d}$
can be estimated from the wavelength where the dust 
emission flux peaks, $\lambda^{\rm d}_{\max}$, and that maximum flux, $F^{\rm d}_{\max}$
\citep{wyatt-2008}.
(Henceforth, any ``flux'' is assumed to be specific flux per unit 
frequency interval.)
If dust behaves as a blackbody, Wien's displacement law gives:
\beq
  T_{\rm d} = 5100\K \left( 1\mum \over \lambda^{\rm d}_{\max} \right) .
\eeq
Assuming, further, that also the stellar photosphere emits as a blackbody,
it is straightforward to find
\beq
  f_{\rm d} =
  { F^{\rm d}_{\max} \over F^{\star}_{\max} }
  { \lambda^{\rm d}_{\max} \over \lambda^\star_{\max} } ,
\label{f_D}
\eeq
where $\lambda^\star_{\max}$ and $F^\star_{\max}$ are
the wavelength where the stellar radiation flux has a maximum and
the maximum flux itself, respectively.
Note that Eq.~(\ref{f_D}) can only be used for ballpark estimates, because
the actual stellar spectrum may deviate from a single blackbody spectrum
considerably.
A more accurate method to determine $f_{\rm d}$ that does not imply
any blackbody assumptions is using its definition:
one calculates the bolometric dust luminosity from the SED, uses the bolometric
luminosity of the star from catalogs, and takes their ratio.


\subsection{Dust Location, Dust Mass, Dust Sizes}
Keeping the assumption that dust interacts with the stellar radiation as a
blackbody and making some assumptions about the disk geometry,
a number of further useful physical parameters can be derived from
$T_{\rm d}$ and $f_{\rm d}$ \citep{wyatt-2008}.
For instance, assuming the disk to be a narrow ring between
$r - {\rm d} r$ and $r + {\rm d} r$,
dust temperature gives the distance from the star:
\beq
   r = \left( 278\K \over T_{\rm d} \right)^2
       \left( L_\star \over L_\odot  \right)^{0.5},
\eeq
whereas $f_{\rm d}$ is directly related to the total 
cross-section area of dust $\sigma_{\rm d}$,
\beq
  \sigma_{\rm d} = 4\pi r^2 f_{\rm d} ,
\eeq
and to the normal geometrical optical depth $\tau_\perp$,
\beq
 \tau_\perp = (r / {\rm d} r ) f_{\rm d} .
\eeq
Dust mass $M_{\rm d}$ can be derived from fractional luminosity as well.
However, this estimation requires further assumptions, namely
about the dust bulk density $\rho_{\rm d}$
and a ``typical''dust size $s_{\rm d}$
(which is poorly defined,
because $f_{\rm d}$-dominating and $M_{\rm d}$-dominating sizes are different).
Then,
\beq
   M_{\rm d} = {16 \over 3} \pi \rho_{\rm d} s_{\rm d} r^ 2 f_{\rm d} .
\eeq

More detailed analyses go beyond the assumptions of blackbody, confinement of dust
to a narrow radial zone, and like-sized dust grains. A usual procedure
\citep[e.g.][]{wolf-hillenbrand-2003} is to assume one or another chemical composition
of dust, as well as a radial distribution of dust from $r_{\rm min}$ to $r_{\rm max}$
and a size distribution from $s_{\rm min}$ to $s_{\rm max}$.
Both distributions are usually postulated to be power laws. Both the limits and 
slopes of these power laws are treated as free parameters, and their values are sought
by fitting the SED and, where avaliable, the resolved images.

\begin{figure}[h!!!]
   \centering
   \includegraphics[width=0.60\textwidth, angle=0]{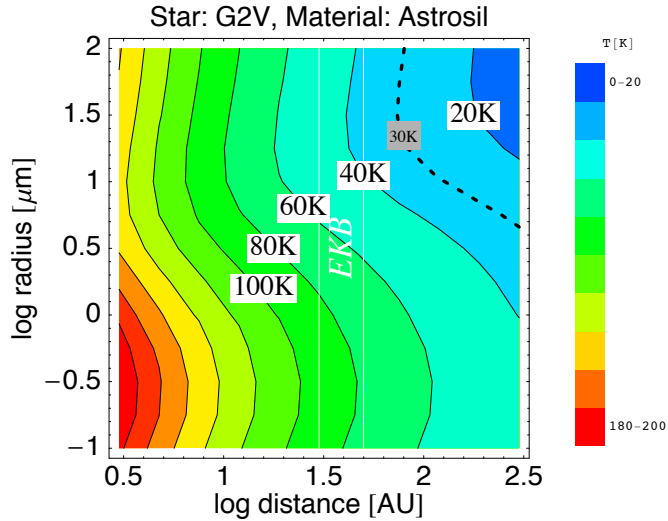}
   \begin{minipage}[]{1.0\textwidth}
   \caption{
   Calculated temperature of different-sized astrosilicate \citep{laor-draine-1993}
   grains at different distances from a G2V star. {\it Courtesy: Sebastian M\"uller.}
   \label{fig:dust_temp}
   }
   \end{minipage}
   \end{figure}

This procedure is very efficient, have been implemented in a number of codes
\citep[e.g.][]{wolf-hillenbrand-2005}
and is intensively used in interpretation of 
the debris disk observations. A short summary of the essential results is given below.
However, it reveals two problems that need to be mentioned.
One is that relaxing the blackbody assumption makes dust temperature dependent not
only of the dust location, but also on the particle sizes
(Fig.~\ref{fig:dust_temp}). Roughly speaking, smaller grains at the same distance
are warmer than larger ones.
This leads to a fundamental degeneracy between distance and size, as the same
SED can be produced equally well by a disk of smaller grains farther out from the star 
and by a disk of larger grains closer in. One way to break this degeneracy is
to invoke information from resolved images, if these are available, as they
show directly where most of the emitting dust is located. Another method is
to invoke theoretical constraints, notably to assume that the minimum grain
size $s_{\rm min}$ is close to the radiation pressure blowout limit $s_{\rm blow}$
(Sect.~\ref{sect:models}). Another problem of the fitting approach described
above is more difficult to cope with. It roots in the assumption that radial
and size distribution are both power laws, independent of each other.
As we saw, this is not what is expected from the debris disk
physics. Instead, one expects the size distribution to strongly depend on distance
and conversely, the radial distribution to be very dissimilar for grains
of different sizes. Additionally, both size and radial distribution can 
substantially depart from single power laws (Fig.~\ref{fig:rad_size_dist}).

\subsection{Outer Disks, Inner disks, and Exozodiacal Clouds}
We now come to a brief overview of essential results on dust location and masses
obtained with the procedures described above.
For many of the systems, the SED is consistent with 
a single, radially narrow dust disk. 
The typical dust temperatures are of the orders of several tens of Kelvin,
which corresponds to the dust disks located at several tens to a hundred AU from the star.
We will refer to them as ``outer disks''.
Such distances readily suggest that the outer disks are likely produced by the 
Edgeworth-Kuiper belt (EKB) analogs around those stars.
However, the typical fractional luminosities of $\sim 10^{-5}$---$10^{-3}$
are at least by two \citep{vitense-et-al-2010} to four 
\citep{booth-et-al-2009} orders of magnitude more than the fractional luminosity
of the presumed EKB dust in the present-day solar system.
The derived dust masses in debris disks (up to $s \sim 1\mm$, best estimated from
sub-mm measurements), are in the range $10^{-3}$--$10^{0}$ Earth mass
\citep[e.g.][]{sheret-et-al-2004,williams-et-al-2004,najita-williams-2005}.

There are also systems that cannot be explained with a single outer dust disk.
Apart from, or instead of, the cold dust disks, some stars have dust 
within $\sim 10\AU$.
The majority of such ``inner disks'' may be associated with collisionally-evolving
asteroid-like belt or sustained by the cometary activity (both
mechanisms are known to be sources of the zodiacal cloud in the solar system).
Alternatively, the fact that many solar-type (FGK) stars show a considerable warm excess
at ages $\ls 100\Myr$, which is about the timescale for accretion of the Earth
in the solar system, suggests that inner dust
may be debris from terrestrial planet formation \citep{kenyon-bromley-2004b}.
The dust masses in the inner debris disks
are estimated to be typically in the range $10^{-8}$--$10^{-6}$ Earth mass
\citep[e.g.][and references therein]{krivov-et-al-2008}.

Yet closer-in to the central stars, within $1\AU$, ``exozodiacal clouds''
have been detected with near-IR interferometry.
The most prominent examples are stars hosting exozodis are
Vega \citep{absil-et-al-2006}, $\tau$~Cet and Fomalhaut 
\citep{difolco-et-al-2007,absil-et-al-2008,akeson-et-al-2009},
as well as $\eta$~Crv \citep{bryden-et-al-2009b}.
The archetypical exozodiacal cloud of Vega has an estimated dust mass of
$\sim 10^{-7}$ Earth mass (equivalent to the mass of an asteroid about $70\km$ in diameter)
and a fractional luminosity of $5 \times 10^{-4}$.

\subsection{Frequencies of Disks around Stars with Different Spectral Types and Ages}
In recent years, detailed statistical studies have been published
that show how the parameters of outer and inner
dust disks depend on the spectral type of the star
as well on the system's age
\citep{meyer-et-al-2004,beichman-et-al-2005,kim-et-al-2005,%
rieke-et-al-2005,bryden-et-al-2006,beichman-et-al-2006b,su-et-al-2006,%
gautier-et-al-2007,siegler-et-al-2006,hillenbrand-et-al-2008,meyer-et-al-2008,%
trilling-et-al-2008,bryden-et-al-2009,carpenter-et-al-2009,plavchan-et-al-2009}.
These statistics are largely based on the Spitzer/MIPS measurements at
$24$ and $70\mum$.
Note that for A-type stars, outer EKB-like disks
are essentially probed at both wavelengths.
But for G-stars, only $70\mum$ excesses correspond to the EKB region while
$24\mum$ measurements detect inner dust inside $\sim 10\AU$
\citep[see, e.g., Fig~1 in][]{wyatt-2008}.
Still longer wavelengths may be required to detect outer disks around K and especially M 
dwarfs, because EKB-sized disks around such stars are very cold.

About 33\% of A-stars over all ages
show excess emission at $24\mum$ and/or $70\mum$, indicative of outer disks 
\citep{su-et-al-2006}.
The disk frequency around A-stars is as high as 60\% at $\sim 10\Myr$ and
declines to about 10\% at $600\Myr$ \citep{siegler-et-al-2006}.
In addition, disks around younger stars are brighter, with a
possible broad peak of the $24\mum$ luminosity at $\sim 10\Myr$ 
\citep{hernandez-et-al-2006}.

For solar-type (F0--K0) stars, the average frequency of outer disk
detections at $70\mum$ was found to be $\approx 16$\% 
\citep{beichman-et-al-2006b,bryden-et-al-2006,trilling-et-al-2008}.
For stars cooler than K1, the fraction may drop to 0\%--4\%
\citep{beichman-et-al-2006b}, and low detection frequencies for K and M
stars have been confirmed by other analyses \citep{gautier-et-al-2007,plavchan-et-al-2009}.
However, as noted above, their detection should
be more efficient at wavelengths longer than $70\mum$.
Indeed, a sub-mm JCMT/SCUBA survey for M dwarfs suggests that the
proportion of debris disks is likely higher than found with Spitzer
\citep{lestrade-et-al-2006}.
The frequencies of disks around FGK stars decline with the age
more slowly that those of A-type stars.
The excess strength, or fractional luminosity,
exhibits a smooth decay with the stellar age as well.

All authors point out a decay of the observed mid- and far-IR excesses
with the system age.
However, the dust fractional luminosity exhibits a large dispersion at any
given age even within a narrow range of spectral classes.
We will come to the interpretation of the long-term decay in debris disks
in Sect.~\ref{sect:planetesimals}.
Apart from the time evolution of the fractional luminosity, attempts have
been made to find correlations between the disk radius on age.
From IRAS 60 and $100\mum$ data,
\citet{rhee-et-al-2007} found that the radii of largest disks increase with
the age, but similar analyses based on $24$ and $70\mum$ Spitzer detections
do not reveal any significant correlation
(\citeauthor{wyatt-et-al-2007b} \citeyear{wyatt-et-al-2007b}
and T. L\"ohne, pers. comm).

Inner disks with a moderate fractional luminosity around A-type stars are
difficult to detect because it requires measurements shortward of $24\mum$
where the photosphere is bright.
Nevertheless, they can be detected through measurements at $24\mum$
around solar-type stars.
The incidences of inner disks drop from 20--40\% at $\sim 20\Myr$ to
only a few percent at $\ga 1\Gyr$ 
\citep{siegler-et-al-2006,meyer-et-al-2008,carpenter-et-al-2009}.
The detection fraction averaged over the ages is $\approx 4$\%
\citep{trilling-et-al-2008}.
Many of older ($\ga 100\Myr$) FGK stars with inner dust have an abnormally
high fractional luminosity, as discussed in  Sect.~\ref{sect:planetesimals}.
Apart from the mid-IR photometry, evidence for inner dust can also be found in the 
Spitzer/IRS spectra in the $\approx 5$ to $35\mum$ range.
\citet{lawler-et-al-2009} analyzed IRS observations of nearby solar-type 
stars and found excess around $12$\% of them
in the long-wavelength IRS band ($30$-–$34\mum$),
but only $1$\% of the stars have detectable excess
in the short wavelength band ($8.5$–-$12\mum$).

Exozodiacal clouds have been found around at least 20\% of AFGKM nearby stars searched 
with IR interferometry (CHARA/FLUOR and Keck/KIN) (Absil et al., in prep.),
There appears to be a weak positive correlation between the incidences of outer
disks and exozodis (O. Absil, pers. comm).

\subsection{Chemical Composition of Dust}
Mid- and far-IR spectra of some debris disks reveal distinctive features
\citep{jura-et-al-2004,chen-et-al-2006,lawler-et-al-2009}, which allows one to get insight
into the mineralogy of the dust grains.
For example, spectra of several disks were matched by a mixture of
amorphous and crystalline silicates, silica, and several other species
\citep{schuetz-et-al-2005,beichman-et-al-2005b,lisse-et-al-2007,%
lisse-et-al-2008,lisse-et-al-2009},
including possibly water ice \citep{chen-et-al-2008}.
The most detailed study was performed for the systems
HD~69830 \citep{beichman-et-al-2005b, lisse-et-al-2007},
HD~113766 \citep{lisse-et-al-2008},
and HD~172555 \citep{lisse-et-al-2009}.
However, interpretation of spectra  is difficult and involves degeneracies,
since the same spectra
can be fitted with different mixtures of materials.
Furthermore, laboratory spectra of various samples used as a reference
have been obtained under conditions (grain sizes and temperature)
that do not necessarily match those in real disks.
Furthermore, the laboratory spectra can differ from one measurement 
technique to another
\citep[e.g., KBr pellets vs. free-floating particles, see][]{tamanai-et-al-2006}.
Despite these caveats, the results are useful. 
Beside the chemical composition, some constraints can be placed on dust
grain sizes and morphology. For instance, the very
presence of distinctive spectral features is indicative of small grains.

\section{Debris disks: Thinking of planetesimals}
\label{sect:planetesimals}

\subsection{Modeling ``from the Sources''}
Results outlined in the previous section describe the ``visible'' dust
component in debris disks,
but they do not tell us directly about the  dust parent bodies and
dust production mechanisms.
Therefore, many of the basic parameters of the debris disk remain obscure.
For instance, while the cross-section area of the disk and thus the observed luminosity
are dominated by small particles at dust sizes,
the bulk of a debris disk's mass is hidden in invisible parent bodies
and cannot be directly constrained from analysis of the dust emission.
Equally, it remains unknown where exactly the planetesimals are located, although
one expects that they orbit the star roughly where most of the dust is seen.
Many properties of the planetesimals, such as their 
dynamical excitation, size distribution, mechanical strength, porosity etc. remain
completely unclear.

There is no direct way to infer
the properties of invisible planetesimal populations
from the observed dust emission. Dust and planetesimals can only be linked
through models.
This is done in two steps.
First, collisional models can be used   
to predict, for a given planetesimal family (mass, location, age, etc.), the 
distribution of dust.
Such models, both analytical and in the form of numerical codes, have become
available in recent years
\citep[e.g.][]{thebault-et-al-2003,krivov-et-al-2006,thebault-augereau-2007,%
wyatt-et-al-2007,loehne-et-al-2007}.
After that, thermal emission models, described in Sect.~\ref{sect:dust},
have to be employed to compute the resulting dust emission.
Comparison of that emission to the one actually observed would then reveal
probable properties of the dust-producing planetesimal families.

\subsection{Constraints from Statistics of Debris Disks}
One way to constrain planetesimal properties with this method is to develop
models for a long-term evolution of disks and to compare their predictions
with the statistics of debris disks of different ages.
As noted above, there is a general tendency of disk dustiness to decay
with time, and this is attributed to the collisional depletion of
planetesimal families.
In an attempt to gain theoretical understanding of the observed evolution,
\citet{dominik-decin-2003} assumed that dust is produced in collisions
among equally-sized ``comets''.
If this dust is removed by collisions, the steady-state amount of dust in such a
system is proportional to the number of comets. This results in an
$M/M_0 \approx \tau/t$
dependence for the amount of dust and for the number of
comets or the total mass of the disk.
Under the assumption of a steady state, this result is valid even for more complex
systems with continuous size distributions from planetesimals to dust.
Tenuous disks, where the lifetime of dust grains is not
limited by collisions but by transport processes like the Poynting-Robertson
drag \citep{artymowicz-1997,krivov-et-al-2000b,wyatt-2005}, should
follow $M \propto t^{-2}$ rather than $M \propto t^{-1}$.
However, slopes as steep as $M \propto t^{-2}$ are not consistent with observations.
Assuming a power-law dependence $t^{-\xi}$,
\citet{greaves-wyatt-2003}
report $\xi \la 0.5$, \citet{liu-et-al-2004} give $0.5 < \xi < 1.0$,
and \citet{greaves-2005} and \citet{moor-et-al-2006} derive $\xi \approx 1.0$.
Fits of the upper envelope of the distribution of luminosities over the age yield
$\xi \approx 1.0$ as well \citep{rieke-et-al-2005}. 
This is an indication that the observed cold debris disks are collision-dominated.

\begin{figure}[h!!!]
   \centering
   \includegraphics[width=0.45\textwidth, angle=0]{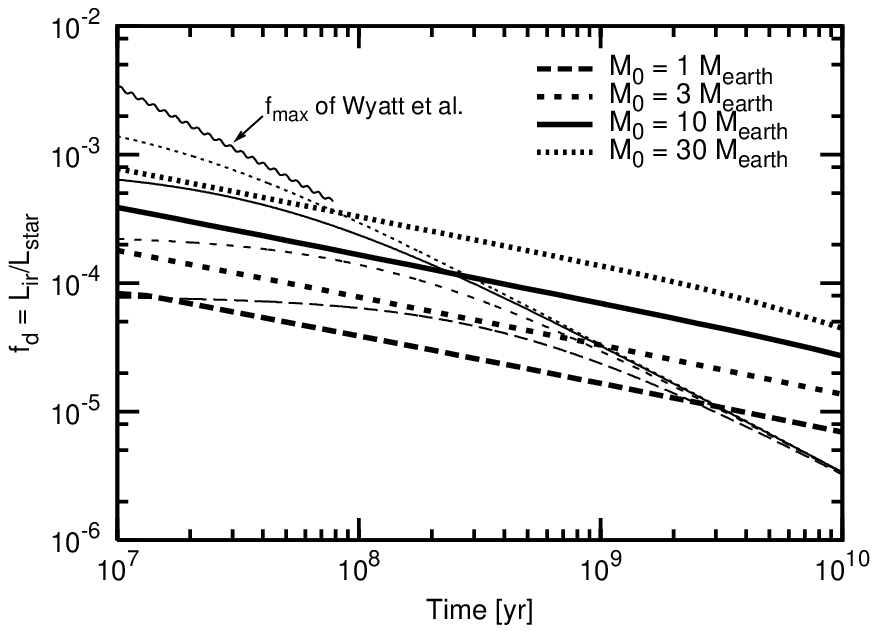}
   \includegraphics[width=0.45\textwidth, angle=0]{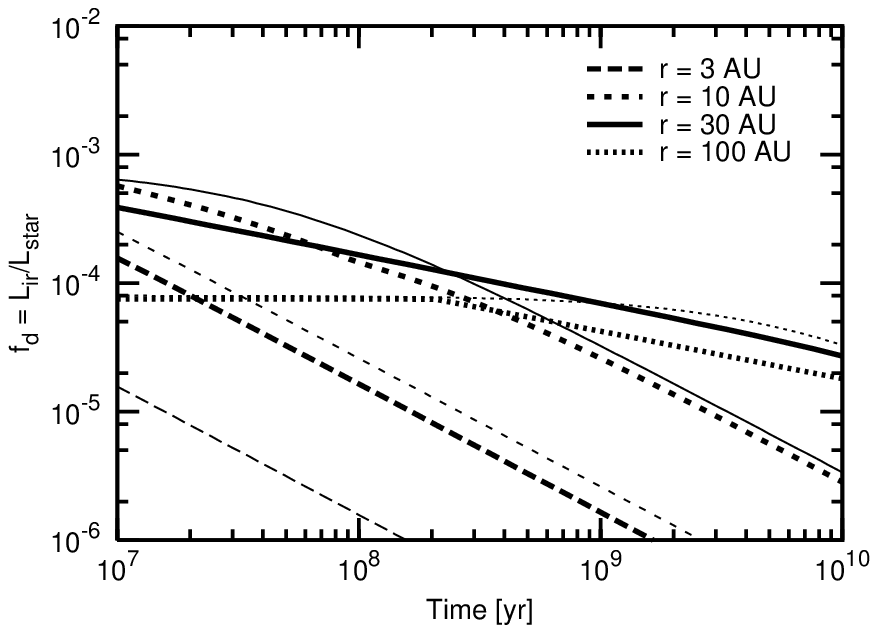}\\
   \includegraphics[width=0.45\textwidth, angle=0]{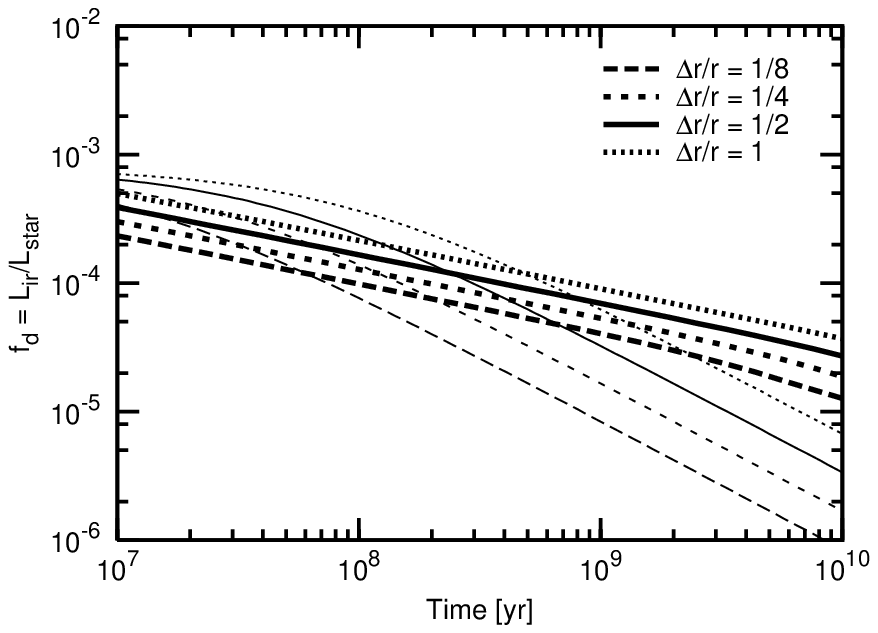}
   \includegraphics[width=0.45\textwidth, angle=0]{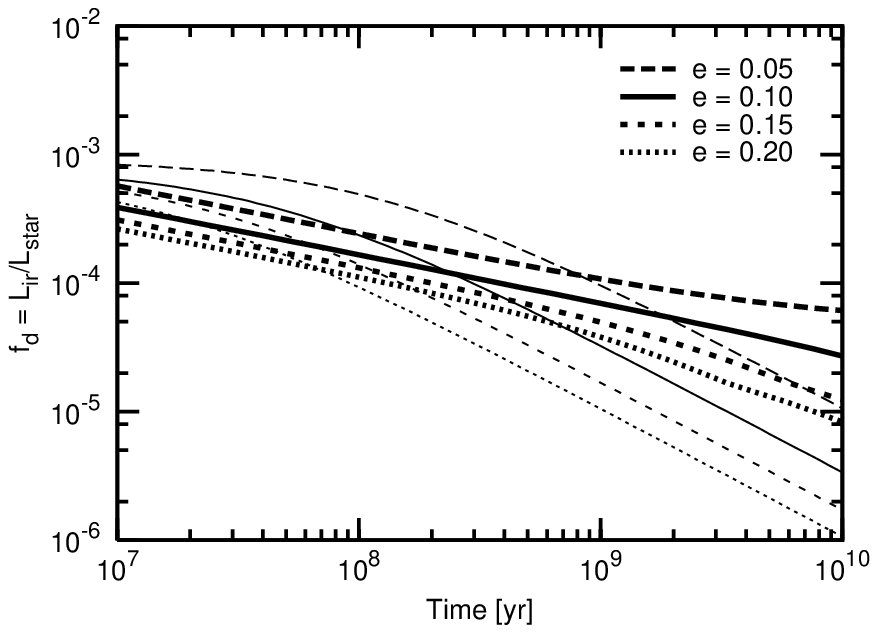}
   \begin{minipage}[]{1.0\textwidth}
   \caption{
  Fractional luminosity of dust around a solar-like star as a function of age.
  Thick lines: analytic model of \citet{loehne-et-al-2007};
  thin lines: Eq.~(20) of \citet{wyatt-et-al-2007}
  with $Q_{\rm D}^*=3 \times 10^4\erg\g^{-1}$ (constant in their model),
  $D_{\rm c} = 60\km$,.
  Different panels demonstrate dependence on different parameters:
  $M_{\rm disk}$ {\it (top left)},
  $r$ {\it (top right)},
  ${\rm d} r/r$ {\it (bottom left)},
  and $e$ {\it (bottom right)}.
  A standard case with
  $M_* = L_* = 1$, $M_{\rm disk} = 10 M_\oplus$,
  $r = 30$~AU, ${\rm d} r/r = 1/2$, and $e = 0.10$ is shown
  with solid lines in all panels.
  {\it Adapted from \citet{loehne-et-al-2007}}.
   \label{fig:long_term}
   }
   \end{minipage}
   \end{figure}

\citet{wyatt-et-al-2007} lifted the most severe simplifying
assumption of the Dominik-Decin model, that of equal-sized parent bodies.
A debris disk they consider is
no longer a two-component system ``comets + dust''. Instead, it is a population of solids 
with a continuous size distribution, from planetesimals down to dust.
A key parameter of the description by \citet{dominik-decin-2003} is the
collisional lifetime of comets, $\tau$.
\citet{wyatt-et-al-2007} replaced it with the lifetime of the largest
planetesimals
and worked out the dependencies on this parameter in great detail.
Since the collisional timescale is inversely proportional to the amount
of material, $\tau \propto 1/M_0$,
the asymptotic disk mass becomes independent of its initial mass
(Fig.~\ref{fig:long_term}, top left).
Only dynamical quantities, i.e. the disk's radial position and extent, the orbiting objects' 
eccentricities
and inclinations, and material properties, i.e. the critical specific energy and the
disruption threshold, as well as the type of the central star determine the very-long-term
evolution (Fig.~\ref{fig:long_term}, thin lines).

Most recently, \citet{loehne-et-al-2007} lifted a major assumption in 
\citet{wyatt-et-al-2007}
that the critical specific energy needed for disruption is constant
across the full range of sizes, from dust
to the largest planetesimals.
As a result, the size distribution of material is no longer a single power law.
One consequence is that at actual ages of debris disks between $\sim$10~Myr and $\sim$10~Gyr,
the decay of the dust mass and the total disk mass follow different laws.
The reason is that, in all conceivable debris disks,
the largest planetesimals have longer collisional lifetimes than the system's age,
and therefore did not have enough time to reach collisional equilibrium.
The mass of visible dust at any instant of time
is determined by planetesimals of intermediate size, whose collisional lifetime is
comparable with the current age of the system, with that ``transitional'' size
gradually increasing with time.
Under standard assumptions, the dust mass, fractional luminosity, and thermal fluxes
all decrease as $t^{-\xi}$
with $\xi = 0.3\ldots 0.4$ 
(cf. Eq.~(\ref{scaling3})).
However, specific decay laws of the total disk mass and the dust mass,
including the value of $\xi$, largely depend on
planetesimal properties, such as
the critical fragmentation energy as a function of size,
the ``primordial'' size distribution of largest planetesimals,
as well as the characteristic eccentricity and inclination of their orbits.

\subsection{Systems with Abnormally High Fractional Luminosity}
Both \citet{wyatt-et-al-2007} and \citet{loehne-et-al-2007} models reproduce reasonably well
the observed statistics of ``cold'' and many ``warm'' debris disks obtained with the
Spitzer/MIPS instrument.
However, there are individual systems that are not consistent with the models.
One implication  of the above models is that they predict a certain maximum fractional 
luminosity $f_{\rm max}$ for a given age that could be produced by a debris disk collisionally
evolving in a steady-state regime.
Assuming physically reasonable initial disk masses of $\la 30 M_\oplus$ 
(corresponding to the mass of solids in the minimum-mass solar nebula),
$f_{\rm max}$ of debris disks around solar-type stars older than $\sim 1\Gyr$
would be as low as $\sim 10^{-4}$ (Fig.~\ref{fig:long_term}, top left).
But some known debris disks have $f \gg f_{\rm max}$.
These are ``warm'' disks within $\sim 10\AU$, exemplified by
the debris disk of an A star $\zeta$~Lep \citep{moerchen-et-al-2007},
as well as some disks of FGK stars:
HD~23514 \citep{rhee-et-al-2008}, $\eta$~Crv \citep{wyatt-et-al-2005},
and HD~69830 \citep{beichman-et-al-2005b, lisse-et-al-2007,payne-et-al-2009}.
The same problem arises for exozodiacal clouds, where $f_{\rm max}$ is quite low
because the collisional evolution as close as at $\sim 1\AU$ from the star
must be very rapid. Thus it is not a surprise that for instance, the exozodi
of Vega \citep{absil-et-al-2006}
appears ``too dusty'' to be explained with a steady-state collisional cascade
in an asteroid belt analog.
The origin of the excessive warm dust and exozodis
is currently a matter of debate 
\citep{wyatt-et-al-2007,wyatt-2008,payne-et-al-2009,wyatt-et-al-2010}.
The observed dust may be debris from recent collisions between two large
planetesimals or planetary embryos \citep{rhee-et-al-2008}.
Alternatively, it can be transported inward from the outer Kuiper belt through
one or another mechanism \citep{wyatt-et-al-2007}.
It is possible that some of the systems with a high amount of warm
dust currently undergo short-lasting, LHB-like instability phase
\citep{wyatt-et-al-2007,booth-et-al-2009}.
Finally, systems like $\eta$~Crv may formally be explained with a collisional cascade in 
a planetesimal swarm with extremely eccentric orbits ($e \sim 0.99$),
although the origin of such population remains challenging
\citep{wyatt-et-al-2010}.

\subsection{Probing Planetesimal Formation Mechanisms}
Since the formation of debris disks is a natural by-product of planetesimal and
planet accumulation processes,
debris disks can help distinguishing between possible planetesimal
formation mechanisms.
The classical scenario is based on a more or less continuous growth of objects from   
micron-sized grains through bigger aggregates that decouple from the gas to,
potentially, planetary embryos. Especially the last stage has been studied to 
predict planetesimal
distributions \citep[e.g.][and previous work]{kenyon-bromley-2008}.
However, theoretical considerations have revealed obstacles
such as the so-called meter barrier. The latter is a ``double bottle-neck'' of
the growth: first, meter-sized objects should be lost to the central star
as a result of gas drag \citep{weidenschilling-1977,brauer-et-al-2007}, and second,
further agglomeration of meter-sized objects upon collision is problematic
\citep{blum-wurm-2008}.
So, in the last years, a competing scenario was invoked
that circumvents these barriers and forms large planetesimals
directly through local
gravitational instability of solids in turbulent gas disks,
which can be triggered, for instance, by transient zones of high pressure
\citep{johansen-et-al-2006} or by streaming instabilities 
\citep{johansen-et-al-2007}.
Both scenarios result in distinctly
different initial conditions for debris disk evolution:
the former predicts a substantial
population of smaller km-sized planetesimals,
while the latter leads to a top-heavy size 
distribution dominated by objects of roughly 100~km in size.
As both scenarios would result in different dustiness of a debris disk at a given age,
there appears to be a principal possibility to distinguish between them
by confronting the model predictions to debris disk observations
(L\"ohne et al., in prep.).

Constraining the top end of the planetesimal size distribution is also important
to understand the mechanism that stirs debris disks.
Extending the formalism of \citet{wyatt-et-al-2007,wyatt-et-al-2007b},
\citet{kennedy-wyatt-2010} devised an analytic model for stirring of debris
disks by the formation of Pluto-sized objects. They showed that the $24$ and $70\mum$
statistics of disks around A-type stars is consistent with this self-stirring
scenario. That also demonstrated that, to satisfy the constraints from observations
such as  the early ``rise and fall'' of peak $24\mum$ excesses around $\sim 10\Myr$
(Sect.~\ref{sect:dust}),
the planetesimal belts must be reasonably narrow 
(${\rm d} r / r \sim 0.5$) and have a minimum radius of $\sim 15\AU$.

\subsection{Constraints from Debris Disks with Known SEDs}
As we could see, useful constraints on the planetesimal properties can be placed
for the statistics of many debris disks, even though observational data on each of them
can be limited to a few photometric points, such as $24$ and $70\mum$ Spitzer/MIPS
measurements. For a limited number of disks, for which photometry points sufficiently
probe the entire SED from mid-IR to sub-mm or mm wavelengths,
additional constraints can be placed.
\citet{krivov-et-al-2008} analyzed five G2V stars with good data (IRAS, ISO/ISOPHOT,
Spitzer/IRAC, /IRS, /MIPS, Keck II/LWS, and JCMT/SCUBA).
For all five systems, the data points could be reproduced
within the error bars with a linear combination of two planetesimal belts
(an ``asteroid belt'' at several $\AU$ and an outer ``Kuiper belt'').
This automatically gave the desired estimates of planetesimals
(location, total mass etc.).
In particular, the cold emission
(with a maximum at the far-IR) is compatible with ``large Kuiper belts'',
with masses in the range 3--50 Earth masses and radii of $100$--$200\AU$.
These large sizes trace back to the facts that the collisional model
predicts the observed emission to stem from micron-sized dust grains,
whose temperatures are well in excess of a blackbody temperature
at a given distance from the star
\citep[as discussed, e.g., in][see also Fig.~\ref{fig:dust_temp}]{hillenbrand-et-al-2008}.
This conclusion is rather robust against variation in parameters
of collisional and thermal emission models, and is roughly consistent
with disk radii revealed in scattered light images
\citep[e.g. HD~107146,][]{ardila-et-al-2004}.
Still, quantitative conclusions about the
mass and location of the planetesimal belts would significantly
depend on (i) the adopted model of collision outcomes
(which, in turn, depend on the dynamical excitation of
the belts, i.e. on orbital eccentricities and inclinations of planetesimals)
and (ii) the assumed grains' absorption and emission efficiencies.
For example, a less efficient cratering
(retaining more grains with radii $\sim10\mum$ in the disk,
see Fig.~\ref{fig:rad_size_dist} left)
and/or more ``transparent'' materials
(making dust grains of the same sizes at the same locations colder)
would result in ``shifting'' the parent belts closer to the star.

   \begin{figure}[h!!!]
   \centering
   \includegraphics[width=0.9\textwidth, angle=0]{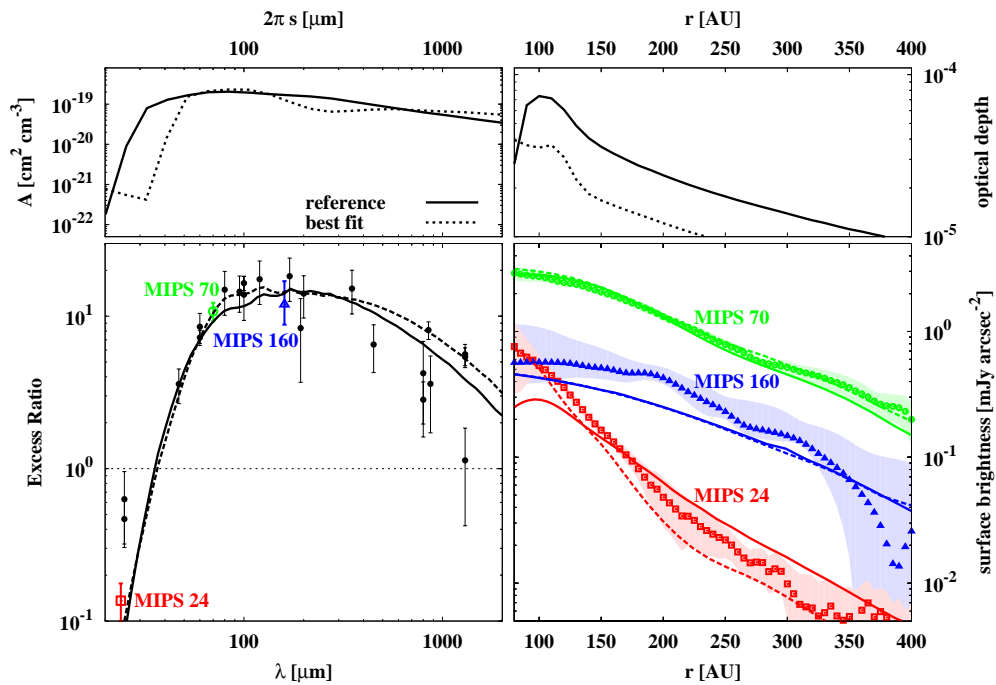}
   \begin{minipage}[]{1.0\textwidth}
   \caption{
     Two models of the Vega disk \citep{mueller-et-al-2009} and their comparison
     with observations.
     The ``reference model'' assumes
     $L_\star = 28 L_\odot$ and a ring of planetesimal with semimajor axes from 80 to $120\AU$ 
     and maximum eccentricities of 0.1.
     The ``best-fit model'' adopts
     $L_\star = 45 L_\odot$, a planetesimal ring from 62 to $120\AU$ with maximum 
     eccentricities of 0.05, and a slightly steeper distribution of fragments in binary
     collisions.
     The reason to vary the stellar luminosity is that Vega is a fast rotator
     \citep{peterson-et-al-2006,aufdenberg-et-al-2006}, which makes stellar 
     parameters functions of the stellar latitude.
     {\it Top left:}
     grain size dust distribution (cross section area density per size decade)
     in the ``reference model''(thick solid line) and the ``best-fit model (dashed line),
     both  in the center of the initial planetesimal ring.
     Note our using $2\pi s$ instead of $s$: since particles with
     the size parameter $2 \pi s / \lambda \sim 1$  emit most efficiently,
     $2\pi s$ roughly gives the typical wavelength of the emission.
     This alleviates comparison between the size distribution and the SED
     (bottom left).
     {\it Top right:}
     radial profile of the normal optical depth for the same disk models.
     {\it Bottom left:}
     corresponding SED in the form of the excess ratio
     (ratio of dust flux to photospheric flux).
     Symbols with error bars are data points
     (large square, large circle, and large triangle mark
      $24\mum$,  $70\mum$, and $160\mum$ excess ratios deduced from
     Spitzer/MIPS images).
     {\it Bottom right:}
     modeled (lines) and observed (symbols, same as in
     the bottom left panel) surface brightness profiles at 24, 70, and $160\mum$.
     The shaded areas around the data points correspond to their error bars.  
     {\it Adapted from \citet{mueller-et-al-2009}}.
   \label{fig:vega_model}
   }
   \end{minipage}
   \end{figure}

\subsection{Constraints from Resolved Debris Disks}
The best constraints of planetesimals and their collisional evolution
can be found from a combined analysis of the SED and resolved images.
Most recently, \citet{mueller-et-al-2009} analyzed
the archetypical debris disk of Vega.
It has been argued before that the resulting photometric data and images may be
in contradiction with a standard, steady-state collisional scenario of
the disk evolution \citep{su-et-al-2005}.
However, \citet{mueller-et-al-2009} 
were able to reproduce the spectral energy distribution in the entire
wavelength range from the near-IR to millimeter, as well as the 
mid-IR and sub-mm radial brightness profiles of the Vega disk
with this scenario (Fig.~\ref{fig:vega_model}).
Thus their results suggest that the Vega disk observations are not in
contradiction with a steady-state collisional dust production, and they put
important constraints on the disk parameters and physical processes that
sustain it.
The total disk mass in $\la 100~\km$-sized bodies is estimated to
be $\sim 10$ Earth masses.
Provided that collisional cascade has been operating over much of the  
Vega age of $\sim 350~\Myr$, the disk must have lost a few Earth masses
of solids during that time.
It has also been demonstrated that using an intermediate luminosity of the star
between the pole and the equator, as derived from its fast rotation, is
required to reproduce the debris disk observations.
Finally, it has been shown that including cratering collisions into the
model is mandatory.

The appearance of the Vega disk in resolved images, with a ring
of large particles seen in sub-mm and extended sheet of small grains
observed in the far- and mid-IR, conforms to the standard debris disk
model described in Sect.~\ref{sect:models}. Some other resolved disks,
such as that of HR8799 \citep{su-et-al-2009}, look similar.
However, the disks of two other A-type stars, Fomalhaut \citep{stapelfeldt-et-al-2004}
and $\beta$~Leo (Stock et al., in prep.), appear narrow at Spitzer/MIPS
wavelengths. The existence of two types of disks~--
wide ones with fuzzy sensitivity-limited outer edges and narrow ones with well-defined
outer boundaries~--- was also concluded by \citet{kalas-et-al-2006}
from scattered-light images.
\citet{thebault-wu-2008} showed that disks with sharp outer edges may arise
from a parent belt of planetesimals with very low eccentricities
($e \sim 0.01$ or less). However, this would imply the absence of planetesimals
larger than several tens of meters, as these would stir the parent belt to 
higher eccentricity values. It is challenging to explain why, for instance,
the protoplanetary disk of Fomalhaut that has built at least one massive planet
at $\sim 100 \AU$ \citep{kalas-et-al-2008},
but failed to produce at least kilometer-sized
planetesimals only slightly farther out from the star.

\section{Debris disks: Thinking of planets}
\label{sect:planets}

\subsection{Structures of Debris Disks as Indicators of Planets}
The usefulness of debris disks goes beyond probing the history of planetesimal 
formation
and planetesimal properties: they can also point to planets.
Indeed, debris disks gravitationally interact with~-- often suspected, but
as yet undetected~-- planets at every moment in their life, and various footprints
of those interactions can be evident in observations of debris dust.
As mentioned above, planets may stir planetesimal disks, launching the
collisional cascade.
Stirring by planets may be
comparable in efficiency with stirring by largest embedded planetesimals
of roughly the Pluto size. Even close-in planets can sufficiently stir
the outer disk, if their orbits are eccentric and the systems are sufficiently
old \citep{wyatt-2005b,mustill-wyatt-2009}.
For instance, the known radial velocity planet of $\epsilon$~Eridani
could have ignited its ``cold'' debris disk
\citep{mustill-wyatt-2009}.

Besides the stirring, a variety of more direct ``fingerprints''
of planets can exist in the disks.
Inner gaps tens of AU in radius commonly observed in resolved debris disk images
\citep[e.g.][and references therein]{moromartin-et-al-2008}
and inferred from the debris disk statistics \citep{kennedy-wyatt-2010}
are likely the
result of clearance by planets \citep{faber-quillen-2007}
(although they may alternatively be ascribed to preferential planetesimal formation 
at larger
distances in some scenarios, \citeauthor{rice-et-al-2006} 
\citeyear{rice-et-al-2006}).
Accordingly, dedicated quests of planets in systems with known ``punched''
debris disks by using direct imaging from space or with adaptive optics
have been initiated in the recent years \citep{apai-et-al-2008}.
The direct imaging method is technically challenging, and only has chances for success
in relatively young systems with
ages younger than a few hundred of Myr, where the planets have not yet
cooled down, but even in that case current searches are not yet sensitive
to planets below about one Jupiter mass.
Despite that, first planets in the inner voids have recently been found around
Fomalhaut \citep{kalas-et-al-2008} and HR~8799 \citep{marois-et-al-2008}.

Another indication for planets could be a
clumpy disk structure observed, for instance, around Vega \citep{wyatt-2003}
and $\epsilon$~Eri \citep{quillen-thorndike-2002}.
Clumps may arise from planetesimal capture into mean motion 
resonances with unseen planets \citep[e.g.][]{wyatt-2006,krivov-et-al-2006b}
or, in systems with a strong P-R or stellar wind drag, from capture
of dust grains drifting inward into such resonances
\citep[e.g.][]{liou-zook-1999,moromartin-malhotra-2002}.
A caveat is that individual breakups of large planetesimals could
sometimes provide an alternative explanation \citep{grigorieva-et-al-2006}.
Attempts to find planets in 
the systems of Vega and $\epsilon$~Eri
by direct imaging have not yet been successful
\citep{itoh-et-al-2006,marengo-et-al-2006,janson-et-al-2007,janson-et-al-2008}.

Warps, such as those observed in the $\beta$~Pic disk,
could be due to secular perturbations by a planet in an inclined orbit
\citep{mouillet-et-al-1997,augereau-et-al-2001}.
In fact, there are many more phenomena in $\beta$~Pic disk that all
point out to at least one Jupiter-mass planet at $\sim 10\AU$
\citep[see][for a summary]{freistetter-et-al-2007}.
The predicted $\beta$~Pic planet  may have already been
detected with VLT/NACO by \citet{lagrange-et-al-2008}, although
this result still remains to be confirmed with a second-epoch
detection \citep{fitzgerald-et-al-2010}.
However, large-scale asymmetries observed in resolved debris disks,
especially the wing asymmetries
in the disks of $\beta$~Pic \citep{mouillet-et-al-1997},
HD~32297 \citep{kalas-2005,fitzgerald-et-al-2007},
and HD~15115 \citep[``the blue needle'',][]{kalas-et-al-2007}
can be attributed to the disk interaction with the ambient interstellar dust
\citep{artymowicz-clampin-1997} or gas \citep{debes-et-al-2009}.

\subsection{Incidences of Debris Disks and Planets around the Same Stars}
Current observational statistics shows that at least $7$\% of
stars searched with radial velocity (RV) method have planets ranging from
super-Jupiters down to super-Earths \citep{udry-santos-2007}\footnote{True
fractions are likely much higher. Most recent analysis of the HARPS GTO survey data
suggests that $30\pm 10$\%  of solar-type stars may host close-in (periods $<$ 100 days),
low-mass ($< 30 M_\oplus$) planets
(Udry et al., paper presented at the ESO/CAUP conference ``Towards Other Earths: perspectives 
and limitations in the ELT era'', Porto, Portugal, 19-–23 October 2009).}.
On the other hand,
IR excesses over the photospheric flux indicative of
(cold) debris disks have been found for about 15\% of main-sequence (MS) stars with 
spectral classes from A to K
\citep{su-et-al-2006,siegler-et-al-2006,trilling-et-al-2008,hillenbrand-et-al-2008}.   
\citet{greaves-et-al-2004a} were apparently the first to raise natural questions:
``do stars with planets also have disks?''
and conversely,  
``do stars with disks also have planets?''.
Answering these questions is hampered by the fact
that the incidences of both planets and disks, retrieved from observational surveys, are 
biased in various ways,   
depend on the current precision or sensitivity of instruments used, and thus
may not reflect the reality.
It is not a surprise, therefore, that the statistical interrelation between the 
presence of 
planets and cold disks remains controversial 
\citep{greaves-et-al-2004a,beichman-et-al-2005,moromartin-et-al-2007,%
greaves-et-al-2006,beichman-et-al-2006b,kospal-et-al-2009,bryden-et-al-2009}.
Nevertheless, it has been found that the incidences of debris disks do not
correlate with the stellar metallicity
\citep{greaves-et-al-2006,beichman-et-al-2005,bryden-et-al-2009},
whereas the typical metallicity of stars with known planets
is well known to be greater than solar \citep[e.g.][]{wright-et-al-2004}.
This might suggest the absence of correlation between the presence of planets
and presence of debris disks, but this conclusion may be premature.
For instance, the incidences of low-mass planets (``Neptunes''), similar
to debris disks, do not seem to correlate with metallicity \citep{sousa-et-al-2008}.
And it is not known if such a correlation exists for more distant planets that
cannot be found with the RV techniques. Thus 
a correlation between the presence of planets and presence of debris disks
cannot be ruled out. In fact, there are
hints for the debris disk harboring systems with a planet detection to possess 
brighter, or dustier, disks  thus higher disk detection rates) than systems without a planet 
detection \citep{beichman-et-al-2005,trilling-et-al-2008,bryden-et-al-2009}.

\section{Debris disks: Thinking of planetary systems}
\label{sect:planetary_systems}

We now focus on those planetary systems that,
like our solar system, are known to contain,
apart the central star and one or more planets,
also an outer and/or inner planetesimal belt, or a combination of them.
A handful of ``full'' systems in this sense have been found so far,
but only first steps have been taken to elucidate their detailed structure.
Examples are
Fomalhaut \citep{kalas-et-al-2008},
$\epsilon$~Eri \citep{backman-et-al-2009},
$q^1$~Eri \citep{liseau-et-al-2008},
HR8799 \citep{reidemeister-et-al-2009},   
HD~38529 \citep{moromartin-et-al-2007b}, and
HD~69830 \citep{beichman-et-al-2005b, lisse-et-al-2007,payne-et-al-2009}.

\begin{figure}[h!!!]
   \centering
   \includegraphics[width=0.45\textwidth, angle=0]{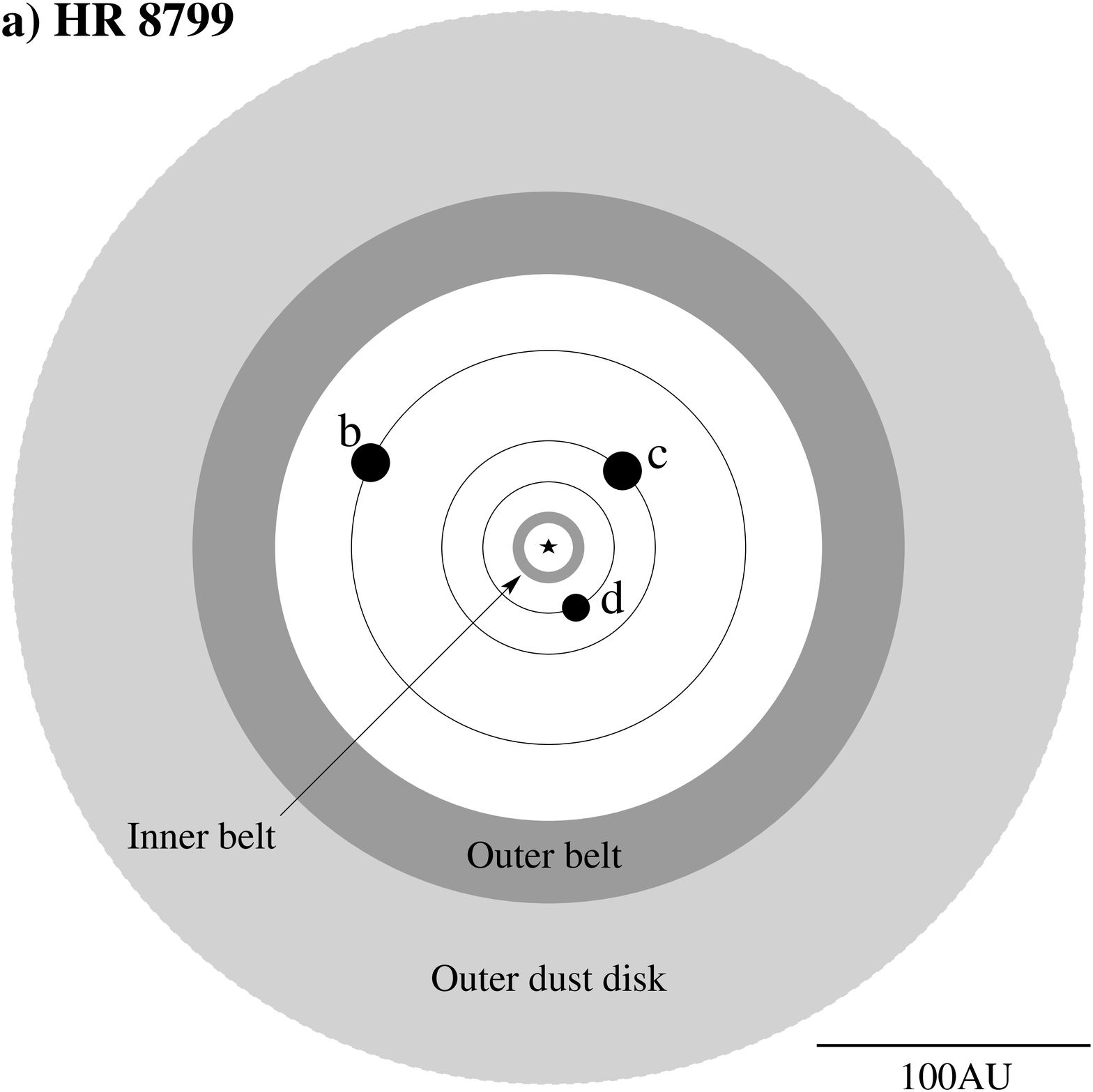}
   \includegraphics[width=0.45\textwidth, angle=0]{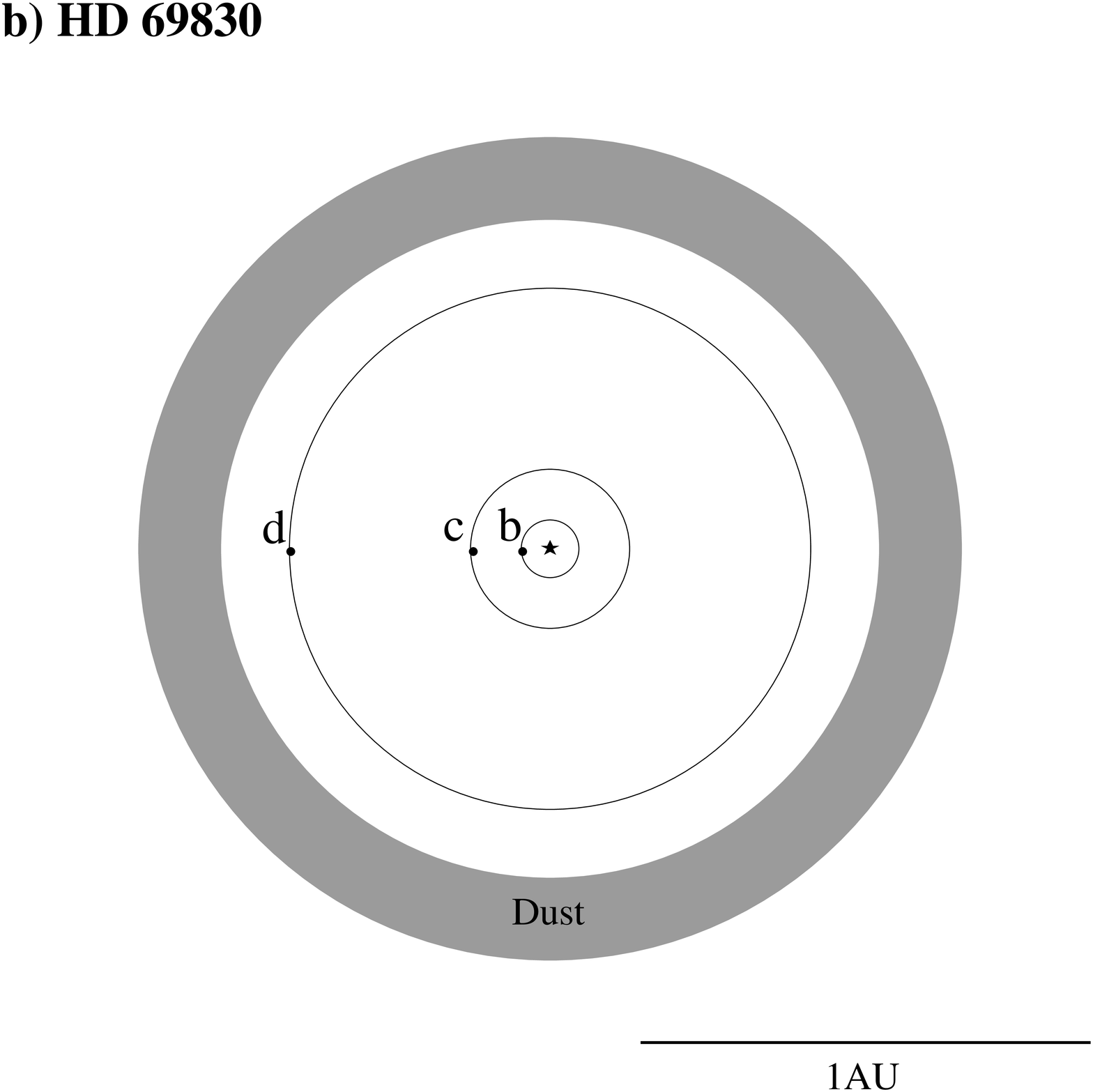}\\
   \includegraphics[width=0.45\textwidth, angle=0]{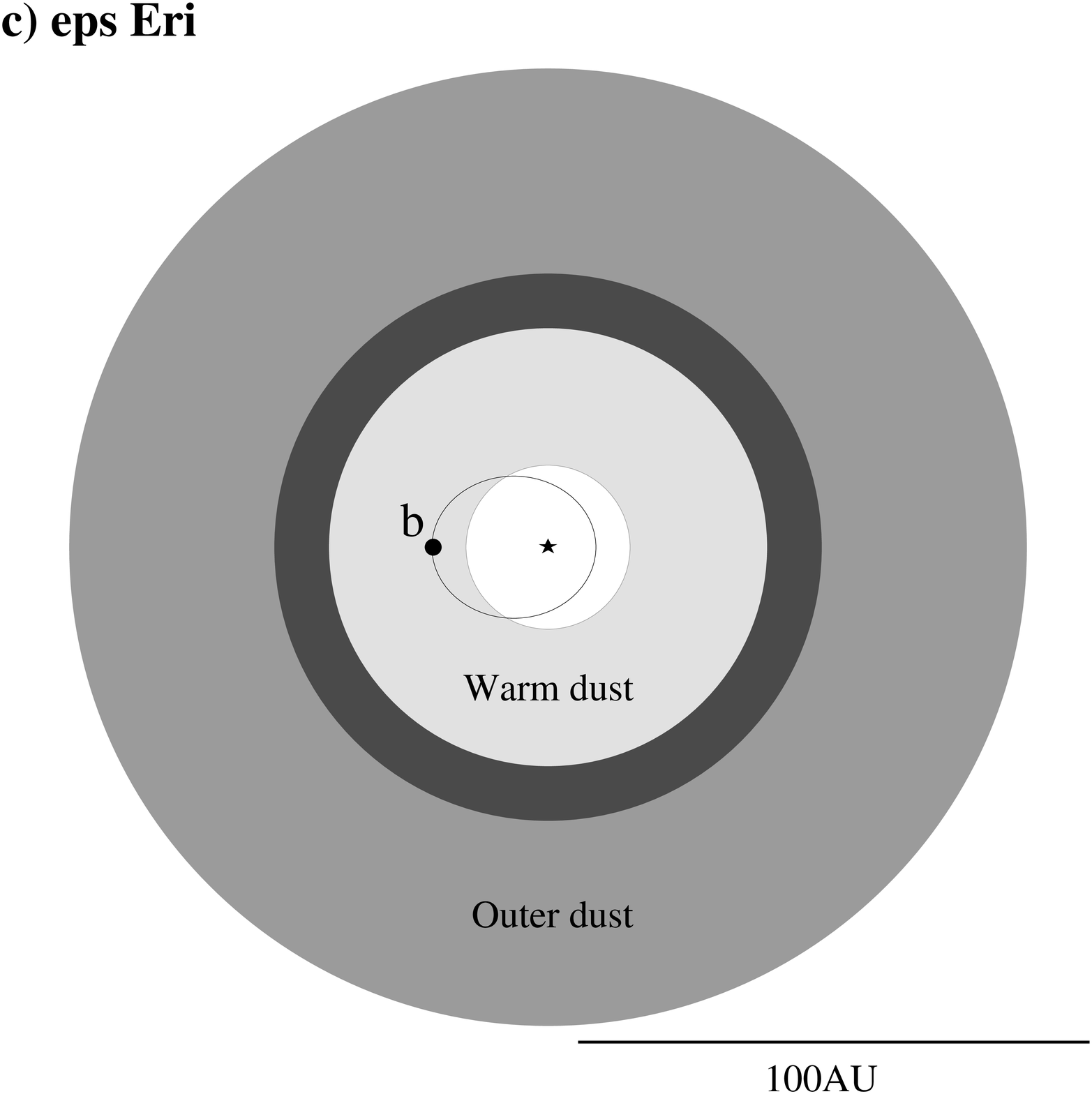}
   \includegraphics[width=0.45\textwidth, angle=0]{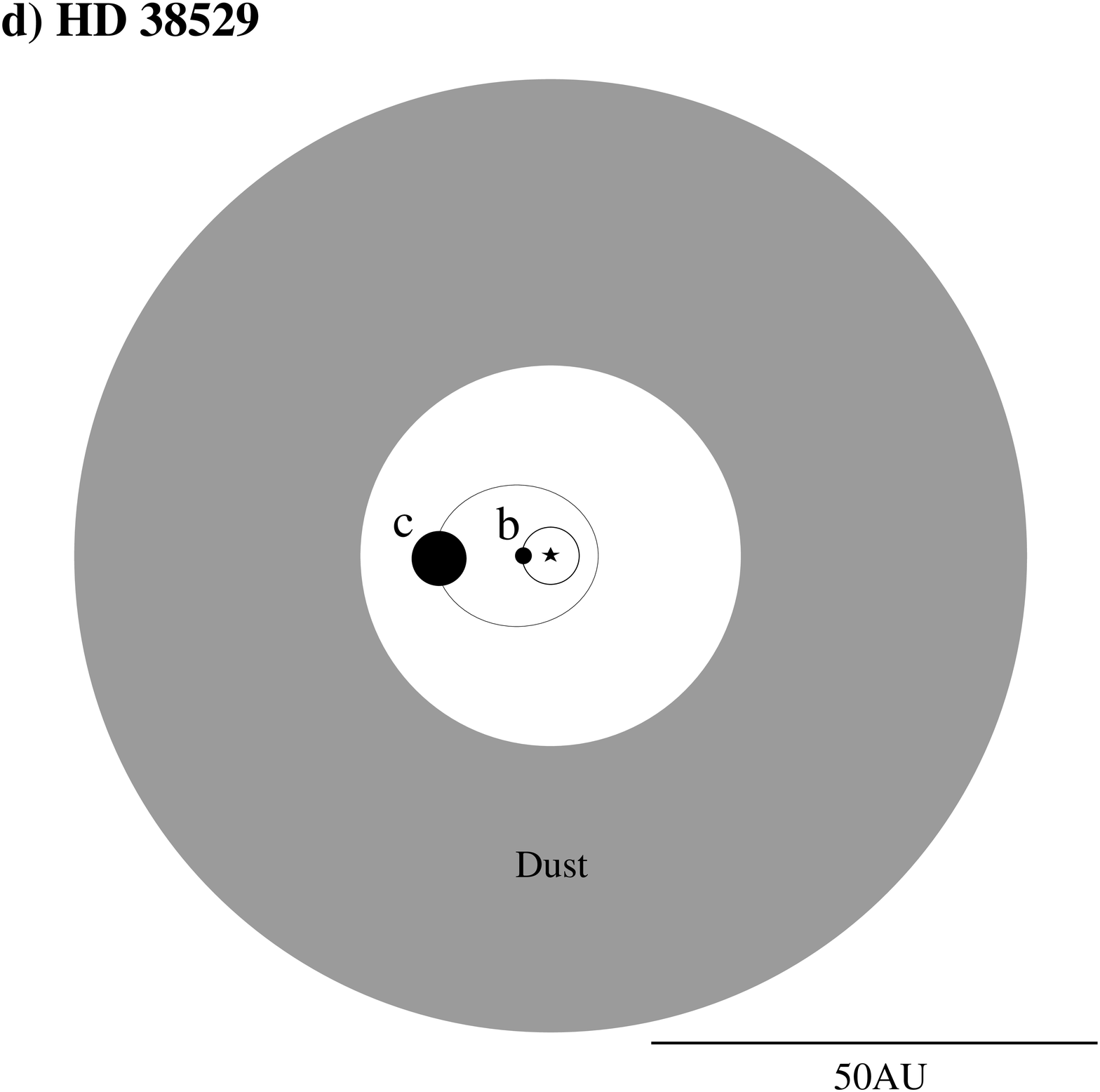}
   \begin{minipage}[]{1.0\textwidth}
   \caption{
   Schematics of four planetary systems with at least one planet and one planetesimal belt:
   {\it (a)} HR~8799, {\it (b)} HD~69830, {\it (c)} $\epsilon$~Eri, and {\it (d)} HD~38529.
   Only known~--- not just presumed~--- components are shown.
   Sketches are not to scale.
   Distance rulers roughly reflect the size of the entire systems,
   the inner parts of the $\epsilon$~Eri and HD~38529 are artificially enlarged.
   The planets are marked with filled circles whose size is representative of their masses.
   \label{fig:sketch_systems}
   }
   \end{minipage}
   \end{figure}

As an example, we take 
the system HR~8799, a nearby A-type star with three directly imaged
planetary candidates \citep{marois-et-al-2008}, of which
the outermost one was recently confirmed by \citet{lafreniere-et-al-2009}
and by \citet{fukagawa-et-al-2009} by analyses of the archival data taken earlier.
Apart from the planets, HR~8799 has long been known to harbor
cold circumstellar dust responsible for excess emission in the far-IR discovered by IRAS
\citep{sadakane-nishida-1986,zuckerman-song-2004,rhee-et-al-2007}.
The rather strong IR excess has also been confirmed with ISO/ISOPHOT
measurements \citep{moor-et-al-2006}, and recently the dust disk was resolved
with Spitzer/MIPS at 24 and $70\mum$ by \citet{su-et-al-2009}.
Additionally,
Spitzer/IRS measurements provided evidence of warm dust emission in the mid-IR
\citep{jura-et-al-2004,chen-et-al-2006}. Both cold and warm dust emission is
indicative of two dust-producing planetesimal belts.
\citet{reidemeister-et-al-2009} undertook 
a coherent analysis of various observational data for all known components of the system, 
including the central star, imaged companions, and dust.
This has led to a view of a complex planetary system with 
an asteroid-like dust producing inner belt at $\sim 10\AU$, three planets of few Jupiter
masses likely locked in a triple Laplace resonance, an outer Kuiper-like planetesimal
belt at $\sim 100\AU$, associated with a tenuous dust disk that extends to  several 
hundreds of AU from the star.
Although many parameters of this system are very different from those
of our solar system (young age, luminous central star, very massive planets),
the arrangement of these components appears similar
(Fig.~\ref{fig:sketch_systems}a).

Other ``full'' planetary systems show less similarity to the solar system.
For instance, the system of HD~69830, a nearby 4-$10\Gyr$ old K0V star, has a very compact 
configuration with three Neptune-mass planets 
within $\sim 1\AU$ \citep{lovis-et-al-2006} 
and a ``warm'' (excess at $8$--$35\mum$) dust belt just exterior of the planetary region 
\citep{beichman-et-al-2005} (Fig.~\ref{fig:sketch_systems}b).
A natural guess that dust can be collisionally produced in an asteroid belt analog
fails, as the dust fractional luminosity is too high
(see Sect.~\ref{sect:planetesimals}).
Several scenarios have been proposed to explain the origin of the warm emission,
but all of them have so far unresolved problems \citep[see][for a detailed 
discussion]{payne-et-al-2009}.

The nearby (3.2~pc), $\la 1$~Gyr old K2~V star $\epsilon$~Eridani
has a ring of cold dust at $\sim 60$~AU seen in resolved sub-mm images
\citep{greaves-et-al-1998, greaves-et-al-2005}, which is encompassed by an extended disk of
warm dust resolved by Spitzer/MIPS \citep{backman-et-al-2009}. The star is orbited by an RV
planet \citep{hatzes-et-al-2000,benedict-et-al-2006,butler-et-al-2006}
with a semimajor axis of 3.4~AU, and another outer
planet is expected to orbit at $\sim 40$~AU, producing the inner cavity and a clumpy structure
in the outer ring \citep{liou-zook-1999, ozernoy-et-al-2000, quillen-thorndike-2002,
deller-maddison-2005}
(Fig.~\ref{fig:sketch_systems}c).
The excess emission at $\lambda \ga 15\mum$ in a
Spitzer/IRS spectrum \citep{backman-et-al-2009} indicates that there is warm dust
at just a few AU from the star.
Its origin is obscure, as an inner ``asteroid belt'' that could produce
this dust would be destroyed by the known inner planet \citep{brogi-et-al-2009}.

One more example of a ``full'' planetary system is the one around
a 3.5 Gyr old G8III/IV star HD~38529 \citep{moromartin-et-al-2007b}.
It hosts two known RV companions
with $M \sin i$ of 0.8 and 12.2 Jupiter masses,
semimajor axes of $0.13\AU$ and $3.74\AU$, 
and eccentricities of $0.25$ and $0.35$, respectively.
Spitzer observations show that HD 38529 has an excess at $70\mum$,
which is compatible with dust coming from a ``Kuiper belt''
between $20$--$50\AU$ (Fig.~\ref{fig:sketch_systems}d).

We do not know yet which combination of planets and dusty planetesimal belts is common,
how these components are typically arranged,
and which circumstances are decisive to set
one or another type of architecture.
However, it is particularly interesting to see how the dust helps to better understand the systems.
We take an example of HR~8799 again.
First, dust helped to narrow the previously reported range of age estimates of the system,
from $30\Myr$ \citep{rhee-et-al-2007} to $\approx 1100\Myr$ 
\citep{song-et-al-2001}.
Namely, the measured IR excess ratio of $\sim 100$ at
$60$~--$90\mum$ would be typical of a debris disk star of age $\la 50$~Myr
\citep[see][their Fig. 5]{su-et-al-2006}. Good age estimates are crucial
from every point of view, but they also improved the
mass estimates of three planets.
Indeed, the masses of three planets are estimated from their observed
luminosities with the aid of the evolutionary models, and the results strongly
depend on the age of the system. 
Old ages would inevitably imply companion masses in the brown dwarf range,
and three companions would then be dynamically unstable,
but for the ages of $\la 50$~Myr there are stable solutions
\citep{gozdziewski-migaszewski-2009,reidemeister-et-al-2009,fabrycky-murrayclay-2009}.
Second, the fact that dust in the outer ring was detected as close as
at 120~AU from the star and that dust in the inner ring extends to at least 10~AU away
from it,  sets independent upper limits of planetary masses that are consistent
with masses from evolutionary models.
Third, dust helped to constrain the orientation of the entire system,
whose knowledge is needed to convert the sky-projected positions into
their true position is space, setting correct initial conditions for the dynamical
stability analyses.
\citet{reidemeister-et-al-2009} showed that
stellar rotational velocity is consistent with an inclination of $13$ -- $30^{\circ}$,
whereas $i \ga 20^\circ$ is needed for the system to be dynamically stable,
thus arguing for a probable inclination of $20$ -- $30^{\circ}$.
That nearly face-on orientation was confirmed by resolved images of the outer
dust disk obtained by \citet{su-et-al-2009}.

\section{Open questions}
\label{sect:conclusions}
After twenty-five years of research on circumstellar debris disks
since the first discoveries around Vega \citep{aumann-et-al-1984}
and $\beta$~Pictoris \citep{smith-terrile-1984}
astronomers can unambiguously place them into the general context
of formation and evolution of planetary systems.
We know that stars form in collapsing interstellar clouds, that young
stars are surrounded by dense, gas- and dust-rich protoplanetary disks
in which planets can form, and that debris disks is what remains around
the star after the dispersal of that primordial gas at ages of
$\sim 10\Myr$ or slightly below.
This view is directly supported by the observations that show a clear
difference between the disks around younger and older stars.
Below those ages, the mass of dust in circumstellar disks retrieved
from the IR excesses is typically $\sim 10^2 M_\oplus$ (Earth mass),
and above them it drops to below one $\sim 1 M_\oplus$.
Applying a standard gas-to-dust ratio of $\sim$~100:1 to protoplanetary disks,
their mass is estimated to be $\sim 10^4 M_\oplus$, which is consistent
with direct gas detections. Direct detection of gas in debris disks is very
difficult, and determination of its mass rather uncertain even where it was detected,
but is it likely that the gas mass does not exceed a fraction of the Earth mass
even in the gas-richest debris disks such as $\beta$~Pic and HD~32297.

One concludes that a certain star possesses a debris disk if
thermal emission of circumstellar dust and/or stellar light scattered by it
has been detected, and if the amount of dust is below a certain limit,
which is taken to be $10^{-2}$ in terms of the fractional luminosity
\citep{lagrange-et-al-2000}.
The latter is required to distinguish between debris disks and protoplanetary disks.
An additional criterion is that the central star should be older than several Myr.
From observations. we can only say that dust is present, and from resolved images
we know that it is present in the form of a disk rather than an envelope, but 
a question is where that dust comes from.
It is easy to estimate the dust lifetimes around a star
against collisions, P-R or stellar wind drag, or destructive processes such as 
photosputtering.
Which removal mechanism is dominant depends on the properties of the central star,
dustiness of the disk, and dust composition~--- but the resulting dust lifetime
is usually well below $1\Myr$. Since debris disks are commonly observed around
stars with ages up to several Gyr, the probability
that we are ``catching'' freshly-born dust is vanishingly low, and so, one or another
mechanism to continually replenish the disks is required. There is currently
little doubt that the main mechanism that supplies dust is collisions between
planetesimals that orbit the debris disk host star. Thus the very existence of debris
disks is a strong evidence for planetesimal populations orbiting a significant
fraction of main sequence stars. This, in turn, tells us that planetesimal
formation in protoplanetary disks is efficient. However, we do not know yet how
these planetesimals typically form. Future debris disk studies may help
discerning between alternative mechanisms of planetesimal formation, since
they predict different size distributions of planetesimals and/or preferential
formation of planetesimals in different regions of protoplanetary disks,
which may be reflected in dust emission at the later, debris disk evolutionary phase.

What is the true incidence of debris disks around main-sequence stars?
Frequency of detection of protoplanetary disks in clusters
younger than $\sim 1\Myr$ is almost 100\% \citep{hernandez-et-al-2007},
but the detection fraction of debris disks is $\approx 15$\% across the spectral stars and 
ages (but $\approx 60$\% for A-stars at $\sim 10$~Myr).
However, this may only reflect the sensitivity of current observations.
It is possible that many other stars, perhaps even 100\%, harbor fainter debris 
disks that as yet remain undetectable.
The same applies to planets. Current RV detections lead to
frequencies of Jupiter-mass planets of $\sim 7$\%, and of lower-mass planets
of $\sim 30$\%, but many more~--- perhaps nearly all~--- stars
may harbor planets that simply escape detection with present-day techniques
and instruments. Statistical relations between systems with planets and
disks remain unclear. We only know that there are stars that have both.
Thus, the typical questions are:
Do nearly all mature stars harbor planets, planetesimals, and dust?
Are there systems that have planets, but
do not have planetesimal belts? Are there systems with planetesimal belts
but without planets? And are there systems, in which protoplanetary
disks failed to produce both?

Accordingly, it is not yet possible to judge about the place 
of our own solar system's planetesimal families, including transneptunian objects,
asteroids, and comets, in the menagerie of debris disks in other planetary systems.
It is clear that the solar system's dust disk, dominated by dust in the Kuiper
belt region, is by orders of magnitude fainter than known debris disks of other
stars. But it is not yet clear why, and, due to current observational limitations,
it is not yet known if any other stars possess disks as tenuous as ours.
And, if they do, what is more common~--- brighter or fainter disks?

We saw that statistics of IR excesses of hundreds of debris disk stars and their evolution
with system's age, as well as observed SEDs and resolved images of some brightest 
debris disks appear consistent with a standard collisional evolution scenario.
It implies one (or more) planetesimal belt with a steady-state cascade of successive
collisions that grinds the planetesimals all the way down to dust, whose
distribution is then shaped by gravity and radiation pressure forces.
In such systems, steady-state collisional models place useful, albeit often weak,
constraints on size and spatial distributions of planetesimals, their dynamical
excitation, and mechanical strength. However, it remains unclear what mechanism
stirs planetesimals to the extent that allows the collisional cascade to operate.
Are these largest, Pluto-sized planetesimals embedded in the disks, neighboring planets,
or both? On the other hand, there are systems~--- or their parts~--- that are certainly not 
consistent with a steady-state collisional evolution. These are exemplified by systems with
substantial ``hot'' dust emission in the mid-IR, such as HD~69830 and $\eta$~Crv,
and by systems with ``exozodiacal'' dust within $\sim 1\AU$ from the star.
Here, there is too much dust to be explained with dust produced locally
in a steady-state collisional cascade. It remains unclear whether non-collisional mechanisms
such as sublimation, one-time major collisions, or non-local production in outer belts
with subsequent inward transport are at work in such systems.

Further open questions are related to the disk structure. Most, if not all, known
debris disks have inner cavities suggesting that planetesimals are not present there.
Nonetheless, it is not known whether the lack of planetesimals is because they originally
fail to form there or were removed by planets later.
Interpretation of azimuthal and vertical structures such as clumps, spirals, warps,
and wing asymmetries remains ambiguous, too. They can be naturally explained
by gravitational interactions of solids with planets, but many can equally well
be attributed to ``non-planetary'' mechanisms such as major collisional break-ups,
stellar flybys, or interaction with the surrounding interstellar gas and dust.

We finally note that detailed studies of thermal emission of debris dust, notably
the analyses of high-resolution mid- and far-IR spectra, offer a unique way
to probe chemical composition of dust and its parent planetesimals. Only the first
steps have been taken in this direction. For instance it is not yet clear, why
some of the disks reveal spectral features and some others do not, to which extent
the inferred composition of dust reflect that of its parent planetesimals,
and whether the dissimilarity of the observed spectra traces back to different
chemical evolution and mixing scenarios at the preceding, protoplanetary disk stage.
Additional information can also be gained by analysis of gas in young debris disks,
which is thought to be a product of dust grain disintegration by various mechanisms.

Future observations and modeling work should bring answers to these and other questions.
On any account, debris disks must be treated as an important constituent
part of planetary systems, and there is little doubt that they 
can serve as tracers of planetesimals and planets and shed light on the planetesimal
and planet formation processes.
Thus the lesson for planetary system research is: ``follow the dust''.

\normalem
\begin{acknowledgements}
This paper is based on the lecture given at
the 1st CPS International School of Planetary Sciences
held in Kobe, Japan, on January 5-9, 2009, and I wish to thank
the organizers for the invitation and support.
This review has further benefited from the author's participation
in the program ``Dynamics of Discs and Planets'', organized
by the Isaac Newton Institute for Mathematical Sciences
in Cambridge.
I thank Torsten L\"ohne, Hiroshi Kobayashi, Sebastian M\"uller,
Martin Reidemeister, and Christian Vitense for stimulating discussions
and Sebastian M\"uller for assistance in preparing the 
figures.
A speedy and careful review by an anonymous referee helped to improve the paper.
Support by the German Research Foundation
(\emph{Deut\-sche For\-schungs\-ge\-mein\-schaft, DFG}), project number
Kr~2164/8--1, by the \emph{Deutscher Akademischer
Austauschdienst (DAAD)}, project D/0707543, and by the International
Space Science Institute in Bern, Switzerland (``Exozodiacal Dust Disks
and Darwin'' working group, http://www.issibern.ch/teams/exodust/)
is acknowledged.
\end{acknowledgements}


\newcommand{\AAp}      {Astron. Astrophys.}
\newcommand{\AApSS}    {AApSS}
\newcommand{\AApT}     {Astron. Astrophys. Trans.}
\newcommand{\AdvSR}    {Adv. Space Res.}
\newcommand{\AJ}       {Astron. J.}
\newcommand{\AN}       {Astron. Nachr.}
\newcommand{\ApJ}      {Astrophys. J.}
\newcommand{\ApJL}      {Astrophys. J. Lett.}
\newcommand{\ApJS}     {Astrophys. J. Suppl.}
\newcommand{\ApSS}     {Astrophys. Space Sci.}
\newcommand{\ARAA}     {Ann. Rev. Astron. Astrophys.}
\newcommand{\ARevEPS}  {Ann. Rev. Earth Planet. Sci.}
\newcommand{\BAAS}     {BAAS}
\newcommand{\CelMech}  {Celest. Mech. Dynam. Astron.}
\newcommand{\EMP}      {Earth, Moon and Planets}
\newcommand{\EPS}      {Earth, Planets and Space}
\newcommand{\GRL}      {Geophys. Res. Lett.}
\newcommand{\JGR}      {J. Geophys. Res.}
\newcommand{\MNRAS}    {MNRAS}
\newcommand{\PASJ}     {PASJ}
\newcommand{\PASP}     {PASP}
\newcommand{\PSS}      {Planet. Space Sci.}
\newcommand{\SolPhys}  {Sol. Phys.}
\newcommand{\SolSysRes}{Sol. Sys. Res.}
\newcommand{\SSR}      {Space Sci. Rev.}


\label{lastpage}

\end{document}